\documentclass[twocolumn]{aastex7}

\definecolor{darkgreen}{rgb}{0.0, 0.5, 0.0}
\newcommand{\commentRA}[1]{}
\begin{document}

\title{J023721.13$-$010528.5: A Giant $S$-shaped Radio Galaxy Showing Four Episodes of Jet Activity}

\correspondingauthor{Dharam V. Lal}
\email{dharam@ncra.tifr.res.in}

\author[0009-0007-4885-5373, sname='Khadekar']{Pavan Vijay Khadekar}
\email{khadekar.pavan@students.iiserpune.ac.in}
\affiliation{Indian Institute of Science Education and Research, Homi Bhabha Road, Pashan, Pune 411008, India}

\author[0000-0001-5470-305X, sname='Lal']{Dharam Vir Lal}
\email{dharam@ncra.tifr.res.in}
\affiliation{National Centre for Radio Astrophysics - Tata Institute of Fundamental Research, Post Bag 3, Ganeshkhind P.O., Pune 411007, India}

\author[0000-0001-7141-7311, sname='Athreya']{Ramana Athreya}
\email{rathreya@iiserpune.ac.in}
\affiliation{Indian Institute of Science Education and Research, Homi Bhabha Road, Pashan, Pune 411008, India}

\begin{abstract}
We present high-sensitivity, high-resolution radio observations of J023721.13$-$010528.5 using the upgraded Giant Metrewave Radio Telescope in band-3 (250--500\,MHz) and band-4 (550--950\,MHz), along with MeerKAT (1280\,MHz) and VLA (3\,GHz) survey images. The radio source, hosted by a bright cluster galaxy at redshift $z=0.372$, is a giant radio galaxy with a projected linear size between the outermost hotspots of 1.2~Mpc and is one of the largest $S$-shaped radio sources known. Its radio morphology revealed four pairs of components straddling the radio core, which were also local maxima in profiles of radio flux densities and equipartition energy density measured along the radio lobes. Spatially resolved multi-frequency spectral analysis showed progressively older emission from the inner to the outer components, spanning approximately 4--21 Myr. The high degree of symmetry of the inner three component-pairs with respect to the radio core makes it improbable that these are randomly located knots in the radio jet. Therefore, we present J023721.13$-$010528.5 as the prototype of a new class of radio galaxies with four episodes of jet activity --- a quadruple-double radio galaxy.
We suggest that giant $S$-shaped radio sources may be among the most promising targets for identifying and studying multi-epoch radio jet activity, since their large linear sizes and jet-axis evolution can spatially separate hotspots from successive episodes, making them easier to detect and analyse.
\end{abstract}

\keywords{\uat{Active galactic nuclei}{16} --- \uat{Giant radio galaxies}{654} --- \uat{Galaxy jets}{601} --- \uat{Radio continuum emission}{1340} --- \uat{Radio galaxies}{1343} --- \uat{Relativistic Jets}{1390}}
%individual: J023721.13-010528.5

\section{Introduction} \label{sec:intro}

Radio jets from powerful radio-emitting active galactic nuclei (AGN), usually termed radio galaxies, are not only interesting in themselves, but are also useful as probes of the environments of the galaxies and the galaxy clusters which host them.

Radio galaxies are commonly classified morphologically as Fanaroff–Riley class I (centre-brightened) or class II (edge-brightened)  \citep{FanaroffRiley1974, Fanaroff2021}.  They are also classified spectroscopically as high-excitation (HERGs) or low-excitation (LERGs) radio galaxies reflecting different accretion modes of the central engine \citep{Hardcastle2020, Mingo2019, Mingo2022, Tadhunter2016}.
The FR\,II radio galaxies are in general more powerful, exhibiting highly relativistic jets of plasma and very luminous terminal hotspots of emission. Some FR\,IIs span many mega-parsecs in size, being the largest structures in the Universe associated with a single cosmic source.

The kilo-parsec scale radio jets powering the hotspots are known to change with time in a small subclass of radio galaxies. The jet changes direction in X-shaped radio galaxies \citep{Merritt2002}, possibly following a merger in a supermassive black hole binary \citep{Lal2004, Lal2006}; though backflow of plasma from the active lobes into older lobes have also been suggested as a possible reason for the observed shape \citep{Capetti2002}. On the other hand, \textit{S}-shaped radio sources suggest a more regular change in the direction of the jet axis, likely due to the precession of the central blackhole \citep{Riley1972, Misra2025}.

Double-double radio galaxies (DDRGs) are another rare subclass of FR\,II radio galaxies characterized by the presence of a second, inner, pair of hotspots \citep{Schoenmaker2000}. This is interpreted as evidence for a second episode of jet activity, with the outer hotspots consisting of old, but still luminous, plasma and the inner ones being the ones powered by the latest active jet. In most of the 192 DDRGs known to date \citep{Nandi2012, Kuzmicz2017, Mahatma2019, Jurlin2020, Dabhade2025}, the two pairs of radio lobes are found to be closely aligned \citep{Konar2006}, suggesting that the jet axis remains largely stable over time, unlike in the \textit{X}- and \textit{S}-shaped sources.

The jets from the first episode of DDRGs are expected to efficiently evacuate the surrounding material along their path, making the formation of new hotspots during the subsequent episode difficult. To explain the formation of the inner lobes in DDRGs, it has been proposed that the surrounding material must refill the evacuated cocoon and increase its density sufficiently for the restarted jets to form new hotspots \citep{Kaiser2000}.
Observationally, DDRGs show flatter spectral indices, younger spectral ages, and higher equipartition magnetic fields for the inner double compared to the outer double \citep{Schoenmaker2000,Konar2006}, consistent with their more recent jet activity.

Similar to DDRGs, an even rarer class of sources is the triple-double radio galaxies (TDRGs), first identified in the source B0925$+$420 \citep{Brocksopp2007}.
Subsequent detections include Speca \citep[also known as J1409$-$0302;][]{Hota2011}, J1216$+$0709 \citep{Singh_2016}, J1225$+$4011 \citep{Chavan2023}, ILTJ145013.45$+$473818.6 \citep{Dabhade2025} and a candidate TDRG J210542.7$-$571912 \cite{Norris2025}, and recently J022248$-$060934 \citep{Rarivoarinoro2026}.  Barring (candidate TDRG) J210542.7$-$571912, the remaining six TDRGs are classified as giant radio galaxies (GRGs), with projected linear sizes exceeding 700\,kpc.
\cite{Hota2011} suggested that the outer lobes in Speca are relic lobes re-energized by external shocks, likely triggered by cluster-scale interactions. ILTJ145013.45$+$473818.6 is hosted by the brightest galaxy in the cluster; \cite{Dabhade2025} have suggested that the asymmetry in its outer lobes is due to the asymmetric environment. \cite{Singh_2016} estimated the kinematic ages of the outer, middle, and inner lobes of J1216$+$0709 to be approximately $1.3 \times 10^8$ years, $7.6 \times 10^6$ years, and $1.5 \times 10^6$ years, respectively, assuming constant jet speeds of 0.01c, 0.05c, and 0.1c. \cite{Brocksopp2007} have reported polarization levels of up to $\sim$20\% in the lobes. \cite{Rarivoarinoro2026} suggested that the triple-double source in their study was a result of short episodes of jet activity followed by long dormant periods.

Multi-epoch radio activity may be difficult to detect in smaller radio sources due to the shorter quiescent phases between jet episodes and the difficulty in separating closely located hotspots from different epochs. Future telescopes with higher resolution and dynamic range (the ability to detect diffuse relic emission against the brighter current hotspot) may result in the detection of multiple events in many more radio sources \citep{Chavan2023}.

Thus, TDRGs constitute an exceptionally rare sub-class of radio galaxies, even in comparison to DDRGs, which are themselves regarded as rare members of this class.  Here, we present a detailed study of the GRG J023721.13$-$010528.5, hosted by an elliptical host galaxy with redshift of 0.37183 in the galaxy cluster WHL J023721.1$-$010528 (see Table~\ref{tab:target}). This remarkable source stands out as one of the rarest of the rare, exhibiting four distinct episodes of jet activity.

Our paper is organized as follows.
The observations and analysis are described in Section 2. Section 3 contains a description of the results from the analysis of optical and radio images, including those related to morphology, spectral indices, and radiative age estimates. In Section 4, we bring together several lines of evidence to conclude that this source indeed shows four distinct episodes of jet activity, making it one of the rarest known examples of recurrent AGN activity, and we discuss its properties in the context of known DDRGs and TDRGs.
Throughout this paper, we adopt $\Lambda$CDM cosmology with $H_0$ = 69.6 km s$^{-1}$ $\mathrm{Mpc}$$^{-1}$, $\Omega_{\rm m}$ = 0.286 and $\Omega_{\Lambda}$ = 0.714. This corresponds to a luminosity distance of 2012.3 $\mathrm{Mpc}$ and a linear scale of about 5.184\,kpc per arcsecond. We define the spectral index, $\alpha$, such that $S_\nu$ $\propto$ $\nu^{\alpha}$, where $S_\nu$ is the flux density at frequency, $\nu$. All positions are in J2000 coordinates.

\begin{deluxetable}{ll}
\tabletypesize{\scriptsize}
\tablewidth{0pt}
\tablecaption{Source Data \label{tab:target}}
\tablehead{
\colhead{Parameter} & \colhead{Value}
}
\startdata
Radio source                                 & J023721.13$-$010528.5 \\
Galaxy cluster                               & WHL J023721.1$-$010528 \\
Optical host galaxy position                 & R.A. 02:37:21.13 \\
                                             & Dec. $-$01:05:28.52 \\
Redshift                                     & 0.37184 \\
Radio Core position $^\dagger$              & R.A.  02:37:21.16 \\
                                             & Dec. $-$01:05:28.60 \\
Projected angular size & 4.2\,arcmin  \\
Projected linear size & outer hotspot separation $\simeq$ 1.2\,Mpc\\
                      & largest linear size $\simeq$ 1.3\,Mpc\\
\enddata

\tablecomments{
$\dagger$ Radio core position from VLASS 3\,GHz image.
}
\end{deluxetable}

\section{Observations and Ancillary Data} \label{sec:Obs}
We observed J023721.13$-$010528.5 with the upgraded Giant Metrewave Radio Telescope~\citep[uGMRT:][]{Swarup1991,Gupta2017} at band-3 (250--500 MHz) and band-4 (550--850 MHz).
We complemented the uGMRT data with the Very Large Array Sky Survey \citep[VLASS:][]{Lacy2025} at 3~GHz single-epoch image of the target. The primary-beam corrected VLASS image has an angular resolution of $\sim$2.5\arcsec\ and a 1$\sigma$ noise of $\sim$140~$\mu$Jy~beam$^{-1}$.
We also used MeerKAT L-band (900--1670\,MHz) data from the MeerKAT Galaxy Cluster Legacy Survey \citep[MGCLS:][]{Knowles2022}.
The corresponding $2^\circ \times 2^\circ$ image has an angular resolution of $\sim$8\arcsec\ and a sensitivity of $\sim$3–5~$\mu$Jy~beam$^{-1}$. Primary beam correction was applied to this MeerKAT image using the \texttt{katbeam}\footnote{https://github.com/ska-sa/katbeam} package. Since J023721.13$-$010528.5 lies near the edge of the MeerKAT primary beam, where the response drops to 20--30\%, the estimated flux density measurements may carry uncertainties of up to $\sim$15\% \citep{de_Villiers2022}.

\subsection{GMRT Observations}

\begin{deluxetable}{lcccc}
\tabletypesize{\scriptsize}
\tablewidth{0pt}
\tablecaption{Log of GMRT Observations}
\label{tab:obsdata}
\tablehead{
\colhead{Observ.} & \colhead{Observ.}    & \colhead{Calibrator} &
                    \colhead{Beam fwhm}  & \colhead{$1\sigma~\mathrm{noise}$} \\
\colhead{date}    & \colhead{freq (MHz)} & \colhead{Flux, Phase} &
                    \colhead{maj, min, PA} & \colhead{($\mu$Jy~b$^{-1}$)}
}
\colnumbers
\startdata
 2023-11-08 & 550--850  & 3C48,      & 7.6\arcsec, 3.5\arcsec, & 30 \\
            & (band-4) & 0323$+$055 & 57.7\arcdeg      &    \\
 2024-05-28 & 300--500  & 3C48,      & 8.7\arcsec, 5.1\arcsec, & 40 \\
            & (band-3) & 0116$-$208 & 46.6\arcdeg      &  \\
            &          & 0323$+$055 &                  & 
\enddata
\tablecomments{Column~4: Synthesized beam parameters --- major-axis, minor-axis, and position angle.}
\end{deluxetable}

GMRT observations (project 45\_063) were carried out on 2023 November 08 in band-4 (550--950 MHz). After two unsuccessful attempts (November 04 and November 27) which were heavily affected by scintillation and radio frequency interference (RFI), we carried out a successful observing run on 2024 May 28 in band-3 (250--500 MHz). We recorded data with 0.67-second integration time and 4096 frequency channels in two polarizations (RR and LL). We analysed only the 300--500 MHz and 550--850 MHz frequency ranges in band-3 and band-4, respectively, as the feed sensitivity drops significantly beyond these ranges.
A flux density calibrator was observed for $\sim$20 minutes at both the beginning and end of each observing session. The observations alternated between the phase calibrator (6 scans of 8 minutes each) and the target source (5 scans of 40 minutes each).
The details of these observations are listed in Table~\ref{tab:obsdata}.

\begin{figure*}
    \begin{center}
        \begin{tabular}{cc}
        \centering
        \includegraphics[width=\linewidth]{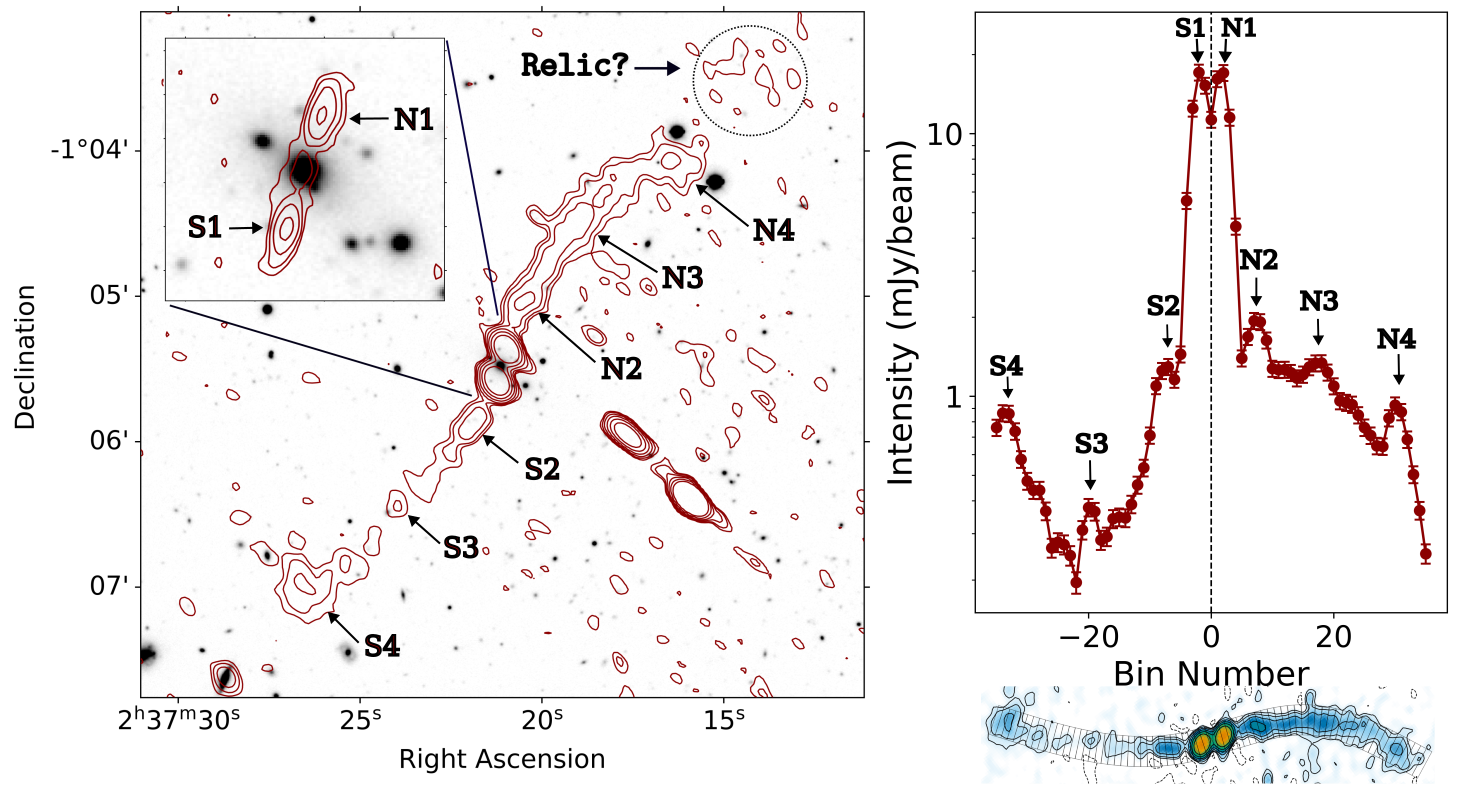}
        \end{tabular}
    \end{center}
    \caption{Radio and optical images of J023721.13$-$010528.5. Left-panel: The grey-scale optical image from the DESI Legacy Survey was obtained by averaging $r$, $g$, $i$, and $z$ band images. The radio contour map is from the uGMRT band-3 (400 MHz) image. The inset shows the high-resolution 3~GHz VLASS image, in which the radio core is well separated from the lobe components and coincident with the host galaxy SDSS J023721.13$-$010528.52. The contour levels are at 3$\sigma\ \times$ [1, 2, 4, 8, 12, 24]. Note the presence of a possible `relic-like' radio emission to the north-west.  The four pairs of components that are the hotspots of four epochs of jet activity are labeled N1 and S1 (north and west of the core, and south and east of the core, respectively; innermost), N2 and S2, N3 and S3, and N4 and S4 (outermost). Right-panel: Peak intensity profile along the ridge-line of the source in the uGMRT 400 MHz image, showing four distinct peaks on both the northern and southern sides of the radio core. The spatial coordinate is represented by the (equispaced) bin number, which is proportional to the projected distance from the radio core. The regions are shown on the radio contour map below the plot. The four pairs of components are associated with local peaks in the intensity profile. The inset is the zoomed-in view of the innermost (N1--S1) pair.}
    \label{fig:optical_bins}
\end{figure*}

\subsection{GMRT Data Reduction}

We manually flagged non-working antennas, as well as spectral channels and time stamps affected by intermittent RFI, in all observations. Calibration and imaging were performed using the Source Peeling and Atmospheric Modeling (\texttt{SPAM}) pipeline \citep{SPAM2014}. \texttt{SPAM} divided the wide-band data into six sub-bands (33.3\,MHz wide in band-3 and 50\,MHz wide in band-4) and processed each sub-band independently as follows.
The flux calibrator was used to set the flux density scale and to perform bandpass calibration, while the phase calibrator provided the amplitude and phase gain solutions. During this stage, the data were also inspected for RFI, bad time stamps, and any other issues, if present. The calibrated target visibilities were then separated into individual data files and processed through a series of imaging and self-calibration steps.
To apply direction-dependent corrections, a point-source sky model was first generated within the primary beam from narrow-band images ($\simeq$ 33.3\,MHz in band-3 and 50\,MHz in band-4) using \texttt{SPAM} and \texttt{PyBDSF}. The full field of view was then divided into facets, and the brightest sources were sequentially peeled. The corresponding direction-dependent phase solutions derived for each peeled source were subsequently applied to the full field. The sub-band image at 800--850\,MHz shows relatively higher noise, likely due to a larger fraction of data being flagged as bad.

The calibrated visibilities from all sub-bands were converted into Measurement Set files and combined into a single data set. A wide-band image was then created using \texttt{WSClean}\footnote{https://gitlab.com/aroffringa/wsclean}, with Briggs parameter, robust = 0, no. of iterations = 50,000, a threshold = 0.5~$\mu$Jy, auto-mask = 3$\sigma$, and a pixel scale = 1.5\arcsec. Primary beam correction was applied using \texttt{wbpbgmrt}\footnote{https://github.com/ruta-k/uGMRTprimarybeam}, and the resulting images were used for all subsequent analysis.
The 1$\sigma$ noise in each image was estimated by averaging values measured from source-free regions around the target source. Images at different frequencies were also re-imaged at a common angular resolution (= 9\arcsec) to produce spectral index maps.
The errors in the estimated flux densities, including calibration and systematic uncertainties added in quadrature, are $\sim$7\% for uGMRT 400-MHz and $\sim$6\% for uGMRT 700-MHz images.

\begin{figure*}
\centering
    \includegraphics[width=\linewidth]{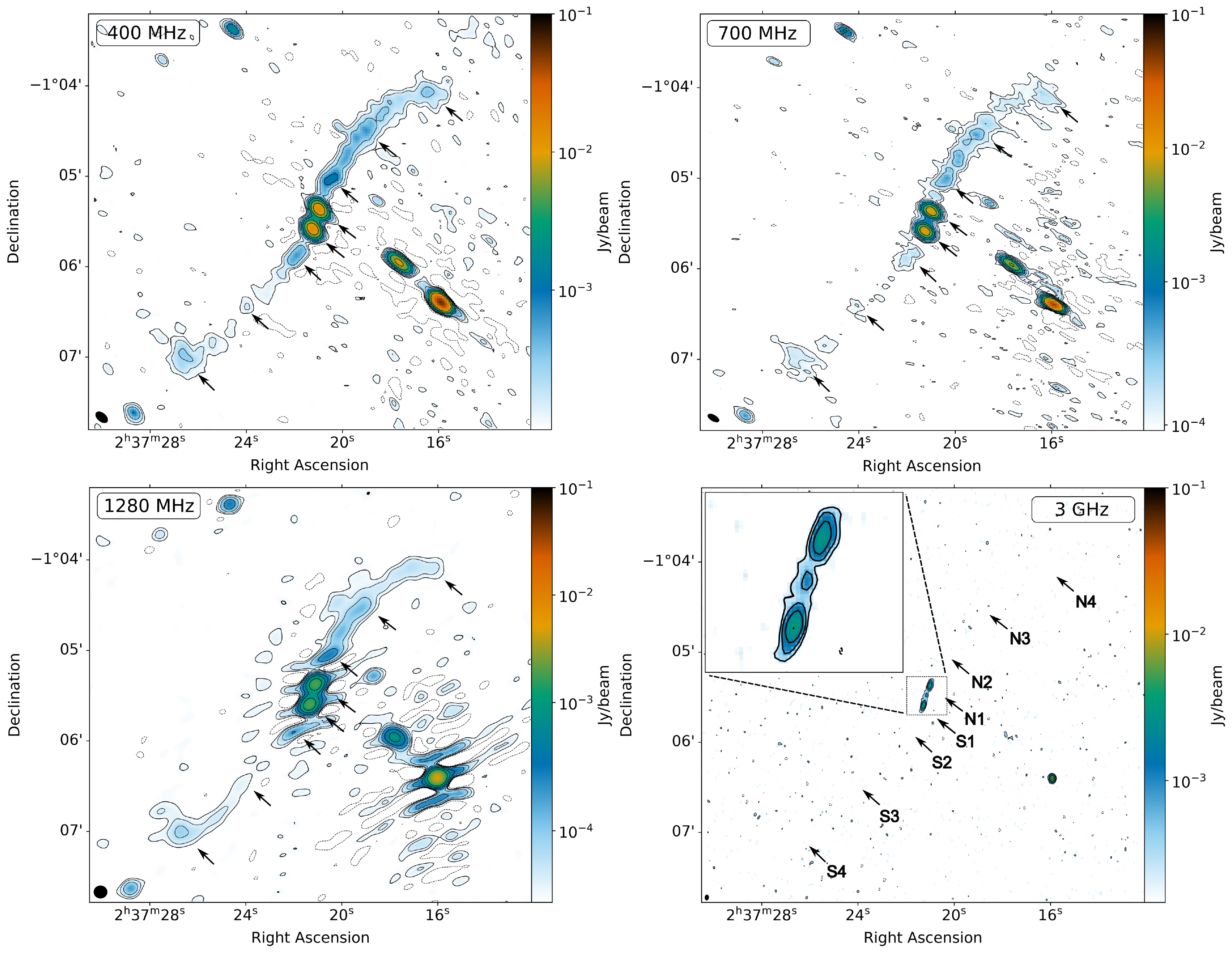}
    \caption{Multi-frequency images of J023721.13$-$010528.5. Top left panel: uGMRT band-3 (400~MHz, resolution = $\sim$8.7\arcsec $\times$ 5.1\arcsec, PA 46.6\arcdeg). Top right: uGMRT band-4 (700 MHz, $\sim$7.6\arcsec $\times$ 3.5\arcsec, PA 57.7\arcdeg). Bottom left: MeerKAT (1280 MHz, $\sim$8.3\arcsec$ \times$ 7.9\arcsec, PA 4.5\arcdeg). Bottom right: VLASS (3 GHz, beam: $\sim$3.1\arcsec$ \times$ 2.1\arcsec, PA $-$3.8\arcdeg); the inset is a zoom of the inner pair of radio lobes and the core. The lowest radio contour is three times the local 1$\sigma$ noise and subsequent contours increase by factors of 2. Arrows mark the four paired compact components in the two radio jets: (left to right) S4--S1 and N1--N4.}
    \label{fig:radio_images}
\end{figure*} 

\begin{deluxetable*}{lcccccccc}
\tabletypesize{\scriptsize}
\tablewidth{0pt}
\tablecaption{Data from Radio Observations \label{tab:fluxdata}  }
\tablehead{
\colhead{Hotspot} &  \colhead{$S_{400~\text{MHz}}$} & \colhead{$S_{700~\text{MHz}}$} &  \colhead{$S_{1280~\text{MHz}}$} & \colhead{$S_{3~\text{GHz}}$} & \colhead{$\alpha$} & \colhead{$\alpha_{\rm low}$} & \colhead{$\alpha_{\rm high}$} & Curvature \\
 & \colhead{(mJy)} & \colhead{ (mJy)} & \colhead{(mJy)} & \colhead{(mJy)} & \colhead{(400-700-1280)} & \colhead{(400-700)} & \colhead{(700-1280)} & \colhead{($\Delta\alpha$)}
}
\colnumbers
\startdata
S4 (outer) & 4.61$\pm$0.37       & 3.34$\pm$0.23    & 1.70$\pm$0.31    &
             $<$3.09$^{\dagger}$ & $-$0.74$\pm$0.14 & $-$0.58$\pm$0.19 &
             $-$1.12$\pm$0.32    & $-$0.54$\pm$0.51 \\
S3         & 1.01$\pm$0.13       & 0.91$\pm$0.10    & 0.42$\pm$0.13    & 
             $<$2.08$^{\dagger}$ & $-$0.49$\pm$0.24 & $-$0.19$\pm$0.30 &
             $-$1.28$\pm$0.54    & $-$1.09$\pm$0.84 \\
S2         & 1.74$\pm$0.14       & 1.26$\pm$0.10    & 0.81$\pm$0.15    &   
             $<$2.00$^{\dagger}$ & $-$0.62$\pm$0.15 & $-$0.58$\pm$0.20 &
             $-$0.73$\pm$0.33    & $-$0.15$\pm$0.53 \\
S1 (inner) & 17.62$\pm$1.24      & 13.71$\pm$0.83   & 9.14$\pm$1.37    & 
             6.18$\pm$0.48       & $-$0.52$\pm$0.12 & $-$0.45$\pm$0.17 &
             $-$0.67$\pm$0.27    & $-$0.22$\pm$0.44 \\ 
\hline
Core       &                     & -                & -                &
             1.39$\pm$0.23       & -                & -                &
             -                   & -                \\
\hline
N1 (inner) & 19.32$\pm$1.36      & 14.90$\pm$0.90   & 8.92$\pm$1.34 &
             6.37$\pm$0.56       & $-$0.59$\pm$0.12 & $-$0.46$\pm$0.17 & 
             $-$0.85$\pm$0.27    & $-$0.39$\pm$0.44 \\
N2         & 4.28$\pm$0.32       & 3.43$\pm$0.22    & 2.00$\pm$0.32 &
            $<$1.96$^{\dagger}$  & $-$0.55$\pm$0.13 & $-$0.40$\pm$0.18 &
            $-$0.89$\pm$0.29     & $-$0.49$\pm$0.47 \\ 
N3         & 5.25$\pm$0.40       & 4.23$\pm$0.27    & 2.07$\pm$0.34 & 
             $<$2.12$^{\dagger}$ & $-$0.63$\pm$0.13 & $-$0.39$\pm$0.18 &
             $-$1.18$\pm$0.29    & $-$0.79$\pm$0.47 \\
N4 (outer) & 5.22$\pm$0.42       & 3.82$\pm$0.27    & 1.63$\pm$0.34 & 
             $<$2.93$^{\dagger}$ & $-$0.79$\pm$0.15 & $-$0.56$\pm$0.19 & 
             $-$1.41$\pm$0.36    & $-$0.85$\pm$0.55 \\ 
 \hline
Entire Source & 61.77$\pm$4.34 & 48.31$\pm$3.82 & 26.58$\pm$4.00 &$<$21.01$^{\dagger}$ & $-$0.64$\pm$0.13 & $-$0.44$\pm$0.19 & $-$0.99$\pm$0.28 &  $-$0.55$\pm$0.47\\
\enddata
\tablecomments{Columns~2-5: Integrated flux densities and 1$\sigma$ error (from Gaussian fitted components) at uGMRT 400 MHz and 700 MHz, MeerKAT 1280 MHz, and VLASS 3 GHz, respectively. \\
Columns 6-8: spectral indices from power-law fit to integrated flux densities across the frequency range (MHz) shown in each. \\
Column 9: spectral curvature is the difference between the high-frequency spectral index and low-frequency spectral index (= $\alpha_{\rm high}$ $-$ $\alpha_{\rm low}$). \\
${\dagger}$: Upper limit on the measured flux density.
}
\end{deluxetable*}

\section{Results} \label{sec:Results}
Figure~\ref{fig:optical_bins} (left-panel) shows the optical image from DESI Legacy Survey overlaid with radio contours from the uGMRT band-3 (400 MHz) image. The inset with the higher-resolution VLASS 3 GHz image \citep{Lacy_2020} shows the positional coincidence between the radio core and the elliptical host galaxy SDSS J023721.13$-$010528.52, a luminous red galaxy with redshift of 0.37183 $\pm$0.00005. The absence of strong emission lines, including H$\alpha$ and [O\,III], indicates that it is a LERG. The velocity dispersion of 317.26 $\pm$16.88 km s$^{-1}$ yields a central black hole mass of 2.28 $\pm$0.71 $\times$10$^{9}\,\mathrm{M_{\odot}}$ \citep{Tremaine_2002, Kormendy2013}, consistent with the higher masses typically found in LERGs with GRGs \citep{Dabhade2020b}.

\subsection{Radio Morphology and Surface Brightness Profile}

Figure~\ref{fig:radio_images} shows images of J023721.13$-$010528.5 at multiple frequencies, including high-resolution, high-sensitivity uGMRT images at band-3 (center frequency = 400 MHz; top left) and band-4 (center frequency = 700 MHz; top right), and ancillary MGCLS (1280 MHz; bottom left) and VLASS (3 GHz; bottom right) images. The MGCLS image shows increased imaging artifacts due to insufficient deconvolution of bright sources south-west of the target source, possibly because the target, J023721.13$-$010528.5 lies at a large angular distance ($\approx 1\arcdeg$) from the phase center of the $2\arcdeg \times 2\arcdeg$ survey image.

The overall radio morphology exhibits an $S$-shaped structure. To study the spatial variation of surface brightness along the source, it was divided into 71 bins (or regions) extending from the north-and-west to the south-and-east with respect to the core, and the peak surface brightness in each bin is plotted in Figure~\ref{fig:optical_bins} (right panel).
Thus, based on this analysis, Figures~\ref{fig:optical_bins} and \ref{fig:radio_images} mark the locations of four pairs of high-surface brightness compact components, labeled N1--N4 (north and west of the core) and S1--S4 (south and east of the core), positioned approximately symmetrically on either side of the host galaxy.
Note that they were identified using a combination of factors, including local peaks in surface brightness (particularly in the uGMRT 400- and 700-MHz images, which have the lowest residual beam sidelobes, see Figure~\ref{fig:optical_bins}), peaks in the equipartition magnetic field profile (see also Section~\ref{spectra-and-curvature}), and a higher degree of compactness, estimated by comparing images at different angular resolutions across the observing bands (see Figure~\ref{fig:radio_images}).
Table~\ref{tab:fluxdata} lists the integrated flux densities and their 1$\sigma$ errors for these four pairs of components.

\subsubsection{Nature of Compact Radio Components}

The innermost pair of components, N1 and S1, are prominently detected in all four frequency bands. The radio core, coincident with the optical host galaxy, is detected as a distinct resolved component only in the high-resolution 3-GHz image. In contrast, the remaining three pairs of components (N2--N4 and S2--S4) are not detected at this frequency.

Among the four pairs, N1 and S1 are the brightest and exhibit the highest degree of symmetry in flux density. The N2 component is approximately 2.5 times brighter than S2, while the asymmetry is even more pronounced for the third pair, with N3 being nearly five times brighter than S3. In contrast, the outermost pair, N4 and S4, have comparable flux densities. The two outermost pairs of components (N3--S3 and N4--S4) also appear progressively more diffuse than the inner pairs.

Two additional components are detected along the source axis beyond N4 and S4. The feature located beyond N4 (toward the north-and-west) is faint but clearly diffuse (S = 2.49 $\pm$0.33 mJy) in the 400~MHz image (see Figure \ref{fig:optical_bins}, left-panel), and appears to be a real ($\approx$ 7.6$\sigma$) extension of the source emission. However, we do not consider it further owing to its non-detection at other frequencies. In contrast, the compact radio emission detected beyond S4 (toward the south-and-east) is associated with an unrelated optical galaxy.

\subsubsection{Radio Spectra and Equipartition Parameters}
\label{spectra-and-curvature}

\begin{figure}
\begin{center}
    \includegraphics[width=\linewidth]{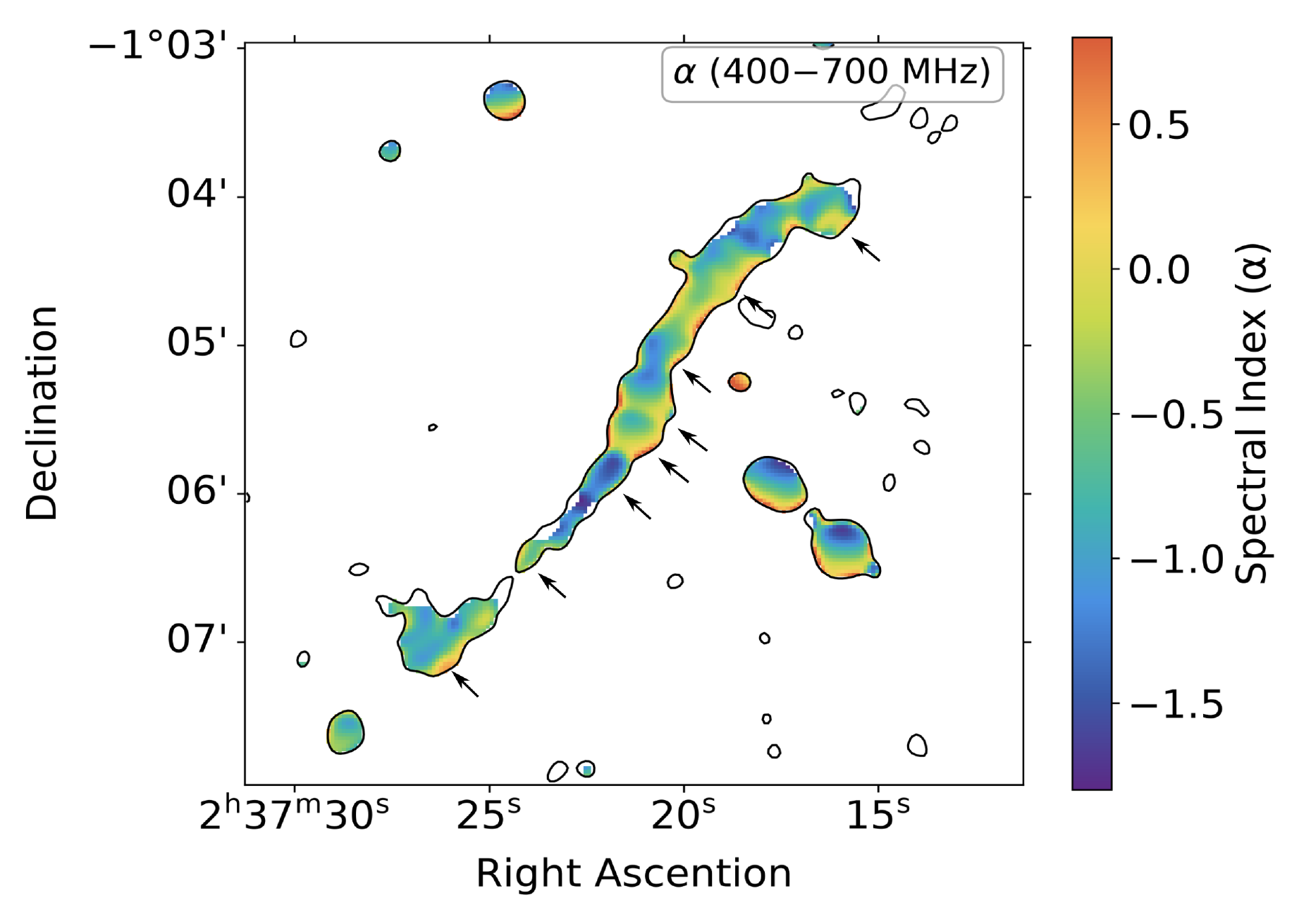}
\end{center}
    \caption{Spectral index map of J023721.13$-$010528.5 from uGMRT 400~MHz and 700~MHz images with matched resolution of 9\arcsec. The outline corresponds to a brightness of 5$\sigma$ noise of the uGMRT 400 MHz image.} 
    \label{fig:spatial_spectral_index}
\end{figure}

\begin{figure}
    \centering
    \includegraphics[width=\linewidth]{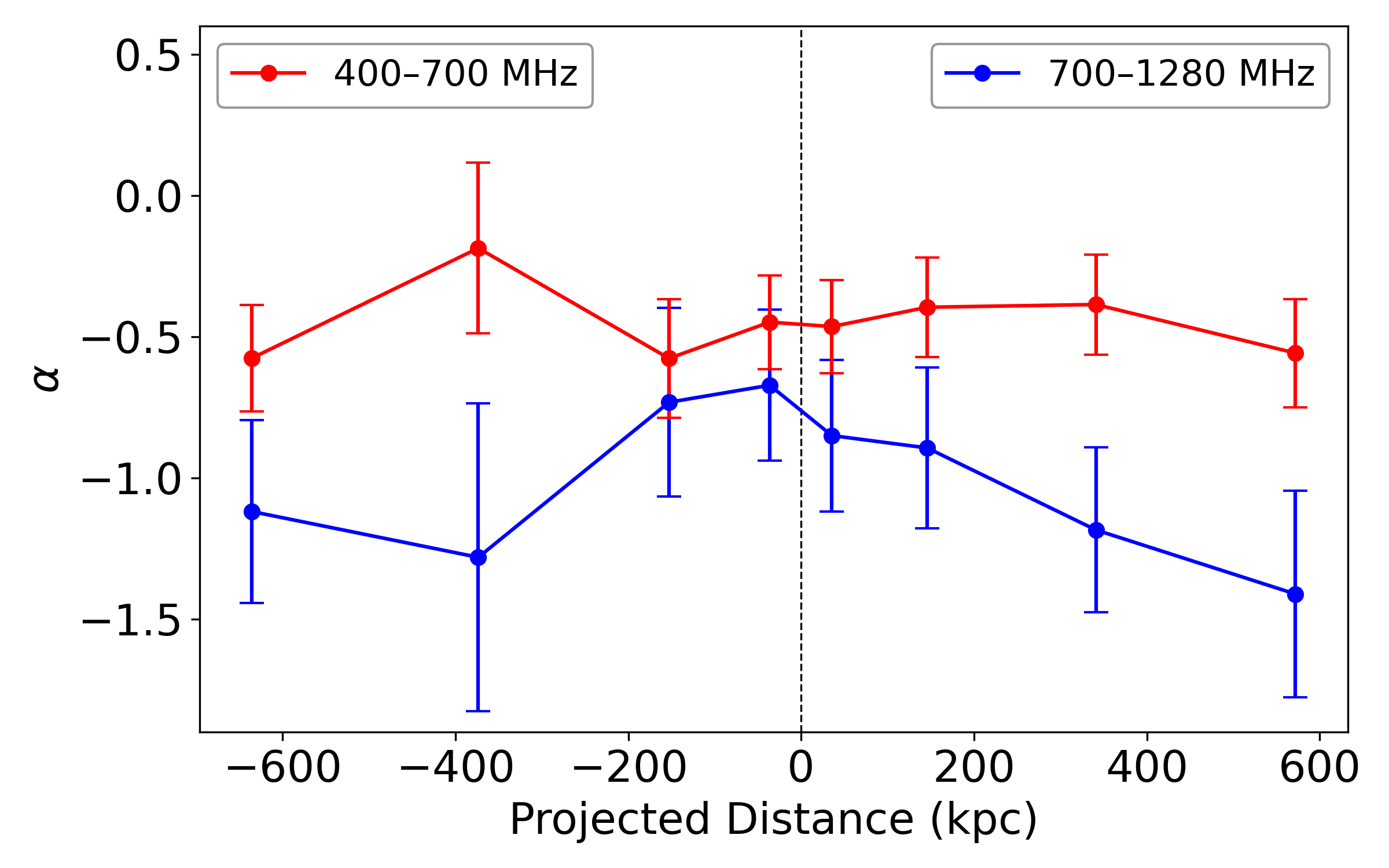}
    \caption{High- and low-frequency spectral indices of the four paired components associated with episodic jet activity. The X-axis shows their distance from the north-and-west to the south-and-east with respect to the core. The dashed vertical line marks the core position.
    }
    \label{fig:alpha_curvature}
\end{figure}

We used 400-MHz and 700-MHz images matched to a common resolution of 9\arcsec\ to construct the low-frequency spectral index map shown in Figure~\ref{fig:spatial_spectral_index}. The map shows a progressive steepening of the radio spectrum with increasing distance from the radio core, with occasional localized flattening at the positions of compact component pairs. 
In order to quantify this, we also derived the low-frequency spectral index using the 400-MHz and 700-MHz images and high-frequency spectral index using the 700-MHz and 1280-MHz images.
The low-frequency and high-frequency spectral indices of the four paired components, derived from their integrated flux densities, are listed in Table~\ref{tab:fluxdata} and shown in Figure~\ref{fig:alpha_curvature}.
The integrated spectra derived from these data (Table~\ref{tab:fluxdata}, Column~6) clearly show a progressive steepening of the radio spectrum with increasing distance from the radio core across all four pairs of compact components.
The high-frequency spectral index profile shows a progressive steepening from the inner component N1 toward the outer component N4 in the north-western jet.  In contrast, the low-frequency spectral index profile is more or less flat across these components within the measurement uncertainties. This trend is less clear in the south-eastern jet due to the lower surface brightness of the radio emission and consequently larger measurement errors, particularly for S3.  

Similarly, the spectral curvature, difference between the high-frequency spectral index and low-frequency spectral index (= $\alpha_{\rm high}$ $-$ $\alpha_{\rm low}$), listed in Table~\ref{tab:fluxdata} (Column~9), shows a similar trend.  It decreases from the inner component N1 to the outer component N4, becoming increasingly curved along the luminous north-western jet. The same trend is less clear in the less luminous south-eastern jet because of the larger uncertainties.

\begin{deluxetable}{llcccc}
\tabletypesize{\scriptsize}
\tablewidth{0pt}
\tablecaption{Equipartition and Model Parameters  \label{tab:eq_values}}
\tablehead{
\colhead{Comp.} & \colhead{$B_{\rm eq}$} & \colhead{$u_{\rm min}$} & \colhead{$P_{\rm min}$} & \colhead{$t_{\rm rad}$} & \colhead{D} \\
 &  & \colhead{($\times 10^{-12}$)} & \colhead{($\times 10^{-12}$)} &  &  \\
 & \colhead{($\mu$G)} & \colhead{(erg cm$^{-3}$)} & \colhead{(dyne cm$^{-2}$)}  & \colhead{(\rm Myr)} & \colhead{($\mathrm{kpc}$)}
}
\colnumbers
\startdata
S4               & 0.6 & 0.03 & 0.01 & 17.5$_{-5.0}^{+4.0}$ & 634.9 \\
S3               & 0.7 & 0.04 & 0.01 & 11.5$_{-11.3}^{+8.0}$ & 372.9 \\
S2               & 0.9 & 0.08 & 0.03 & 13.5$_{-12.7}^{+8.5}$ & 140.4 \\
S1               & 1.3 & 0.15 & 0.05 &  4.5$_{-4.5}^{+4.0}$ &  35.0 \\
\hline
N1               & 1.3 & 0.15 & 0.05 &  7.5$_{-5.0}^{+3.5}$ &  35.3 \\
N2               & 1.0 & 0.08 & 0.03 & 11.2$_{-10.8}^{+7.3}$ & 146.1 \\
N3               & 0.9 & 0.07 & 0.02 & 17.5$_{-6.0}^{+5.0}$ & 338.4 \\
N4               & 0.6 & 0.03 & 0.01 & 20.5$_{-4.5}^{+4.0}$ & 562.7 \\
\hline
% Core$^{\dagger}$ & 3.8 & 1.33 & 0.44 &    - &    -  \\
Source    & 1.3 & 0.15 & 0.05 &    - &    -  \\
\enddata
\tablecomments{
Columns~2--4: equipartition magnetic field, minimum energy density, and equipartition pressure \citep{Beck2005}.
Column~5: Radiative age \citep[JP model:][see also Section~\ref{sec:spectral-age}]{Beck2005,JP}.
Column~6: Projected linear distance from the radio core.
}
\end{deluxetable}

\begin{figure}
    \centering
    \includegraphics[width=\linewidth]{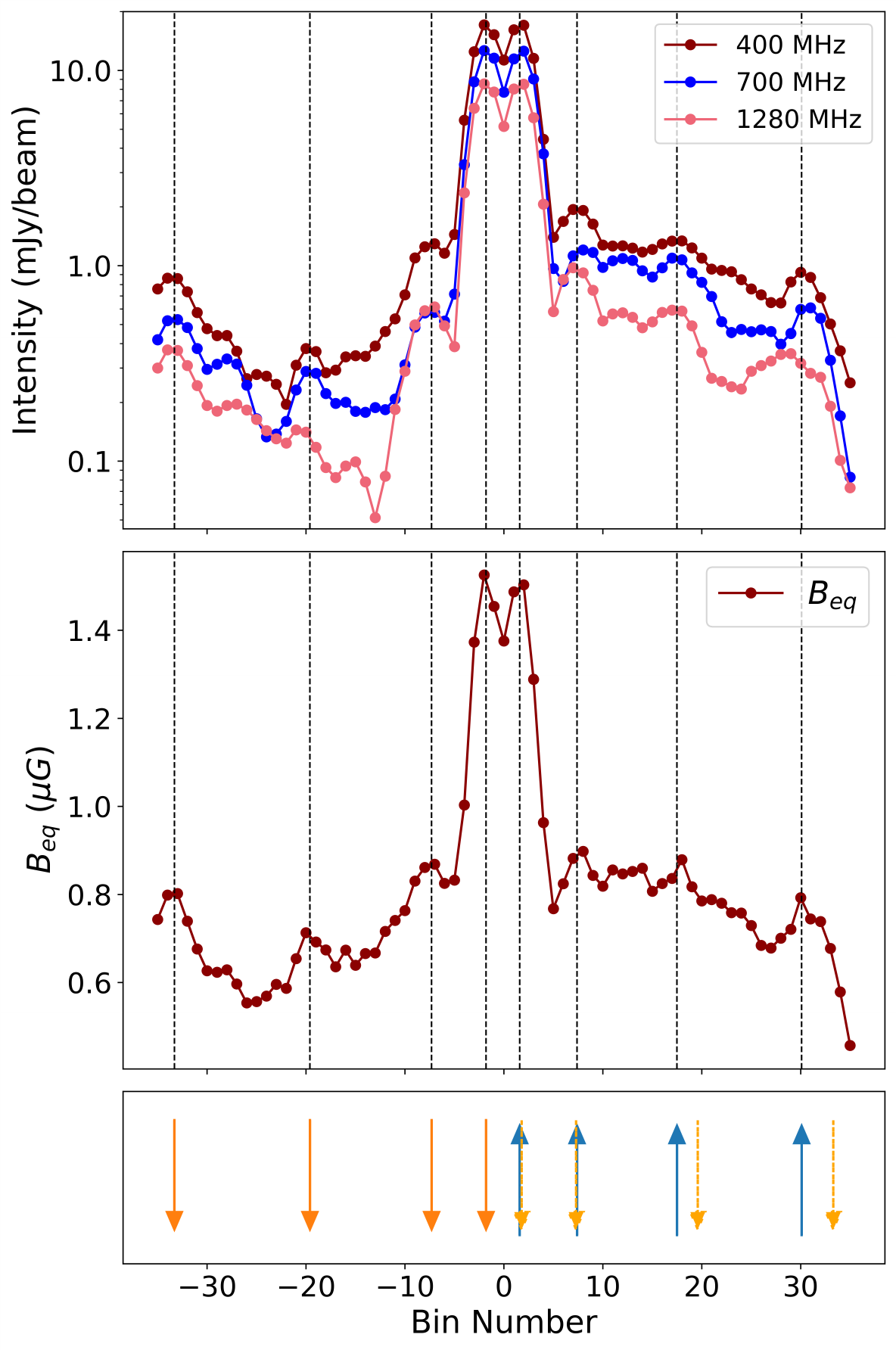}
    \caption{Top panel: Multi-frequency comparison of the peak intensity profile along the source ridge-line. The bin number represents the spatial coordinate along the source (also see Figure~\ref{fig:optical_bins}).
    Middle panel: Profile of the equipartition magnetic field.
    The locations of four pairs of compact components are shown as vertical dotted lines.
    Lower panel: locations of the four paired components (from left to right) S4--S1 (left half: downward orange arrows) and N1--N4 (right half: upward blue arrows). The reflection of S4--S1 in the right half (downward yellow arrows) illustrates the degree of symmetry between the four pairs of compact components. }
    \label{fig:eqipartition}
\end{figure}

We investigated the energetics of different radio components and their evolution across successive episodes
of jet activity, by estimating the average magnetic field strength and the total energy in each component (Table~\ref{tab:eq_values}) using the formalism of \cite{Miley1980} and \citep{Beck2005}. In this calculation, we assumed equipartition between relativistic particles and the magnetic field, equal energy density in heavy particles and electrons, a unit filling factor for the emitting regions, a line-of-sight path length equal to the projected transverse size of each component, a spectral index of $\alpha=-0.8$ over the frequency range 0.01--100~GHz, and a minimum Lorentz factor $\gamma_{\rm min}=100$.
These assumptions, along with the measured flux densities (Table~\ref{tab:fluxdata}, Columns~2--4), provide the derived physical conditions for all regions of the source. The resulting equipartition parameters (and equipartition pressure) for the source and four pairs of components are listed in Table~\ref{tab:eq_values}.
Figure~\ref{fig:eqipartition} shows the variation of surface brightness (upper panel), equipartition magnetic field (and equipartition pressure, middle panel) across the source. The regions with enhanced equipartition magnetic field (and equipartition pressure) broadly coincide with the peaks in the surface brightness profile, i.e., the locations of four pairs of compact components (see Figure~\ref{fig:eqipartition}, lower panel), which further supports the identification of the component peaks, including the very faint S3 component.  The magnetic field strength (and equipartition pressure) shows a systematic decrease from the innermost (N1--S1) to the outermost (N4--S4) component pairs.
% The highest equipartition magnetic field is observed in the core ($\sim$3.8~$\mu$G). 

\subsection{Kinematic and Spectral Ages}
\label{sec:spectral-age} 

Even though the radio lobe is curved it is reasonable to assume that successive hotspots propagated along a straight-line connecting their current position and the AGN core, and estimated the duration of jet activity for each episode, expressed as a kinematic age.

The projected separations between the hotspots corresponding to different episodes of jet activity are $\sim$70 kpc (N1--S1 pair), $\sim$286 kpc (N2–S2 pair), $\sim$711 kpc (N3–S3 pair), and $\sim$1198 kpc (N4–S4; outermost pair).  Assuming a typical hotspot advance speed of 0.1c, consistent with $\sim0.01$–0.15c inferred for FR\,II radio galaxies \citep{Arshakian2000, Kaiser1997}, we derive kinematic ages of $\sim$1.1, $\sim$4.7, $\sim$11.6, and $\sim$19.5 Myr, respectively.
We note that these kinematic ages represent the duration of jet activity in each episode, and not the temporal sequence in which the episodes were triggered.

\begin{figure}[t]
    \centering
    \begin{tabular}{cc}
    \includegraphics[height=14.7cm]{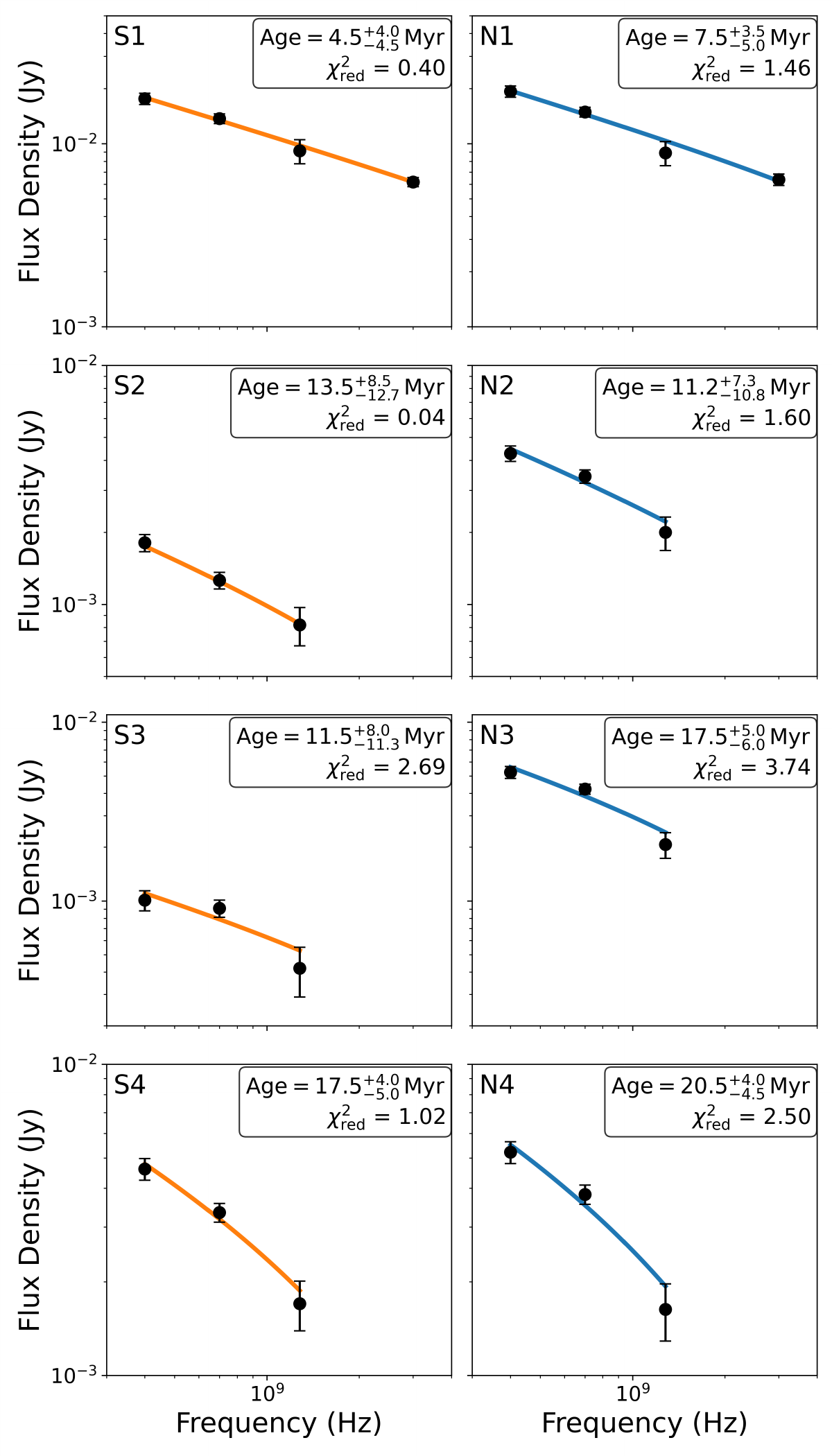}
    \end{tabular}
    \caption{Integrated spectra and spectral age distribution \citep[JP model;][]{Beck2005,JP} of the four principal pairs of components (hotspots; N1--N4 and S1--S4). The spectral plots include the model age estimate and goodness of fit estimated from the integrated flux density of the compact region in the vicinity of the hotposts.}
    \label{fig:ages_inner}
\end{figure}

We derived the spectral ages of the four paired components using the uGMRT and MeerKAT data by fitting several models including the JP \citep{JP}, KP \citep{Kardashev1962, Pacholczyk1970}, JP-Tribble, and KP-Tribble \citep{Tribble1993} models in the \texttt{BRATS} package \citep{BRATS}. We have only reported the JP values since the others were consistent with it. We assumed an electron injection index $\alpha_{inj}$ = $-$0.5, a continuously isotropic electron pitch-angle distribution in a uniform magnetic field, a filling factor of the emitting regions unity, and a path length through the component along the line of sight equal to its transverse size. The field strength was obtained from equipartition analysis (see Section~\ref{spectra-and-curvature}). We selected a circular region around the peak of each component to average over multiple pixels to estimate the spectral age with better accuracy. For the inner components, we used a 3$\sigma$ mask to define the region since the coincidence of the peaks across frequencies was quite good. This coincidence was less pronounced in the more diffuse outer components, so we averaged the measurements over a larger area for them.

The spectral age errors provided by the BRATS package are asymmetrical about their best age estimate, which is difficult to use in any subsequent regression analyses. So, we calculated BRATS spectral ages of 1000 Monte Carlo (MC) simulations using the observed flux densities modulated by their respective uncertainties. The median radiative ages and errors are listed in Table~\ref{tab:eq_values}). The fits to the integrated spectra are shown in Figure~\ref{fig:ages_inner}.

The resulting spectral ages span $\sim$4.5–20.5 Myr. Since the north-western and south-eastern jets are expected to be active simultaneously, the corresponding pairs of compact components should have similar radiative ages. Hence, any difference in the derived ages within a component pair provides an estimate of the random uncertainty in the spectral-age determination. The age differences between paired components range from $-$2.3 to +6 Myr, with a standard deviation of 4.5 Myr which was calculated assuming that the mean difference between the north-south components was zero, as is appropriate for co-eval hotspots. As expected, the largest discrepancy is associated with the very faint S3 component with respect to its counterpart, N3.

\begin{figure}
\centering
\includegraphics[width=\linewidth]{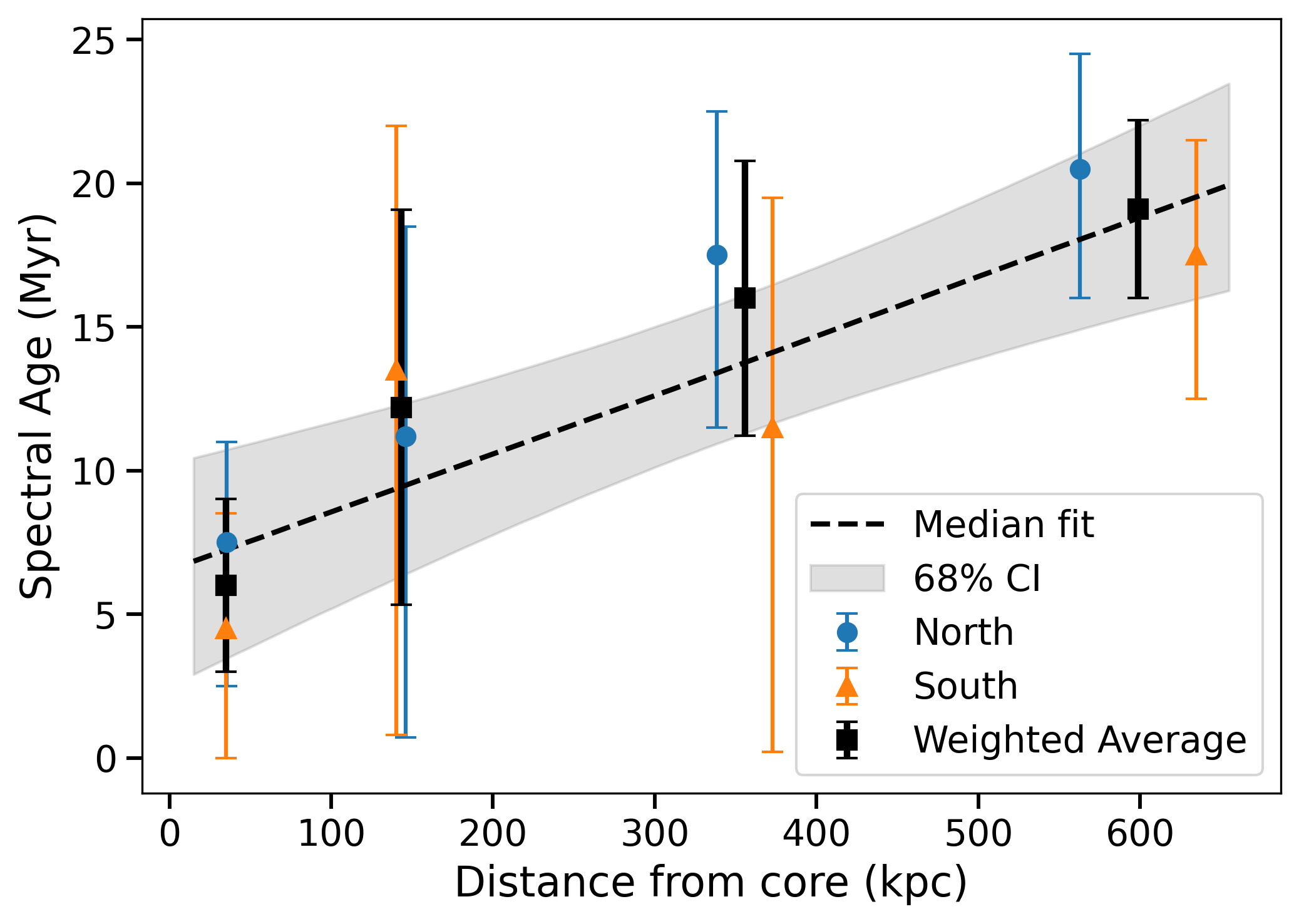}
\caption{Spectral ages derived from the integrated spectra of individual lobes associated with episodic jet activity are plotted as a function of their projected distance (in kpc) from the radio core. The black dashed line shows the straight-line fit from the MC simulations, while the grey shaded region represents its $\pm$1$\sigma$ dispersion (see Section~\ref{sec:spectral-age}).}
\label{fig:age_dist_plot}
\end{figure}

Figure~\ref{fig:age_dist_plot} shows the variation of radiative age with projected distance from the radio core, along with the straight-line fit and its $\pm$1$\sigma$ dispersion. The dispersion was estimated by separately fitting each of the aforementioned Monte-Carlo data sets.

Both jets exhibit a general increase in radiative age with distance from the core, a trend that is particularly clear in the north-western jet owing to its higher surface brightness and consequently smaller age uncertainties.  A similar trend was obtained when the radiative age of each compact component is estimated as the mean of the distribution of radiative ages of the individual pixels within that component using \texttt{BRATS} \citep{BRATS}.

%%%%%%%%%%%%%%%%%%%%%%%%%%%%%%%%%%%%%%%%%%%%%%%%%%%%%%%%%%%%
%%%%%%%%%%%%%%%%%%%% FIGURE ... spectral age vs dist

\section{Discussion}\label{sec:Discussion}

The radio source J023721.13$-$010528.5 is identified with the brightest cluster galaxy, SDSS J023721.13$-$010528.52 \citep{Yuan2016}, located at the center of the galaxy cluster WHL J023721.1$–$010528. The cluster has a richness of 14.49, a spectroscopic redshift of 0.3718, and 11 members within its $R_{200}$ radius of 0.95 Mpc \citep{Wen2015}.

\subsection{Knots in the Jet or Multi-epoch Jet Activity?}
\label{sec:multiepoch}

As a compact component (or `hotspot') advances, it evacuates a tunnel through the circumgalactic medium~\citep{Morganti2014}, reducing the drag on the jet~\citep{Lal2013} and thereby producing little jet emission between the core and the hotspot (although the lobe plasma, which is the backflow from the hotspot, itself may remain luminous). However, if a dense cloud of plasma is not fully dispersed by the passing hotspot, it can give rise to knots of emission well before the hotspot. We present two lines of evidence, based on spectral age and symmetry arguments, to distinguish such knots from multi-epoch hotspots.
\begin{enumerate}
    \item If these features were knots within a single, continuously active jet, their synchrotron-emitting plasma would display similar radiative ages because they are energized by the same jet moving at relativistic speeds --- the time difference between first and fourth component should be less than the travel time of the jet ($=$ 2 Myr). In a multi-epoch scenario, the hotspot pair being currently powered by the jet should have the least radiative age, while the other older pairs are aging passively.
    \item If the six observed components (located inside the two outermost hotspots) were produced by encounters of the jet with randomly located dense clouds (i) their numbers on either side of the core need not be the same, and (ii) even when paired across the radio core they are not constrained to be symmetrically located. In contrast, hotspots produced in each jet episode should form in (approximately) symmetric pairs on opposite sides of the nucleus.
\end{enumerate}

Figures~\ref{fig:ages_inner} and \ref{fig:age_dist_plot} show the integrated spectra (flux density as a function of frequency) and the model-derived spectral ages as a function of the projected straight-line distance (in kpc) from the radio core for the four principal pairs of compact components. They indicate that the spectral ages of the four hotspot pairs span a range (4.5 to 20.5 Myr) which is considerably larger than the spectral age error estimate of 4.5 Myr derived from the difference in values of north-south component pairs.

\begin{figure}
    \centering
    \includegraphics[width=\linewidth]{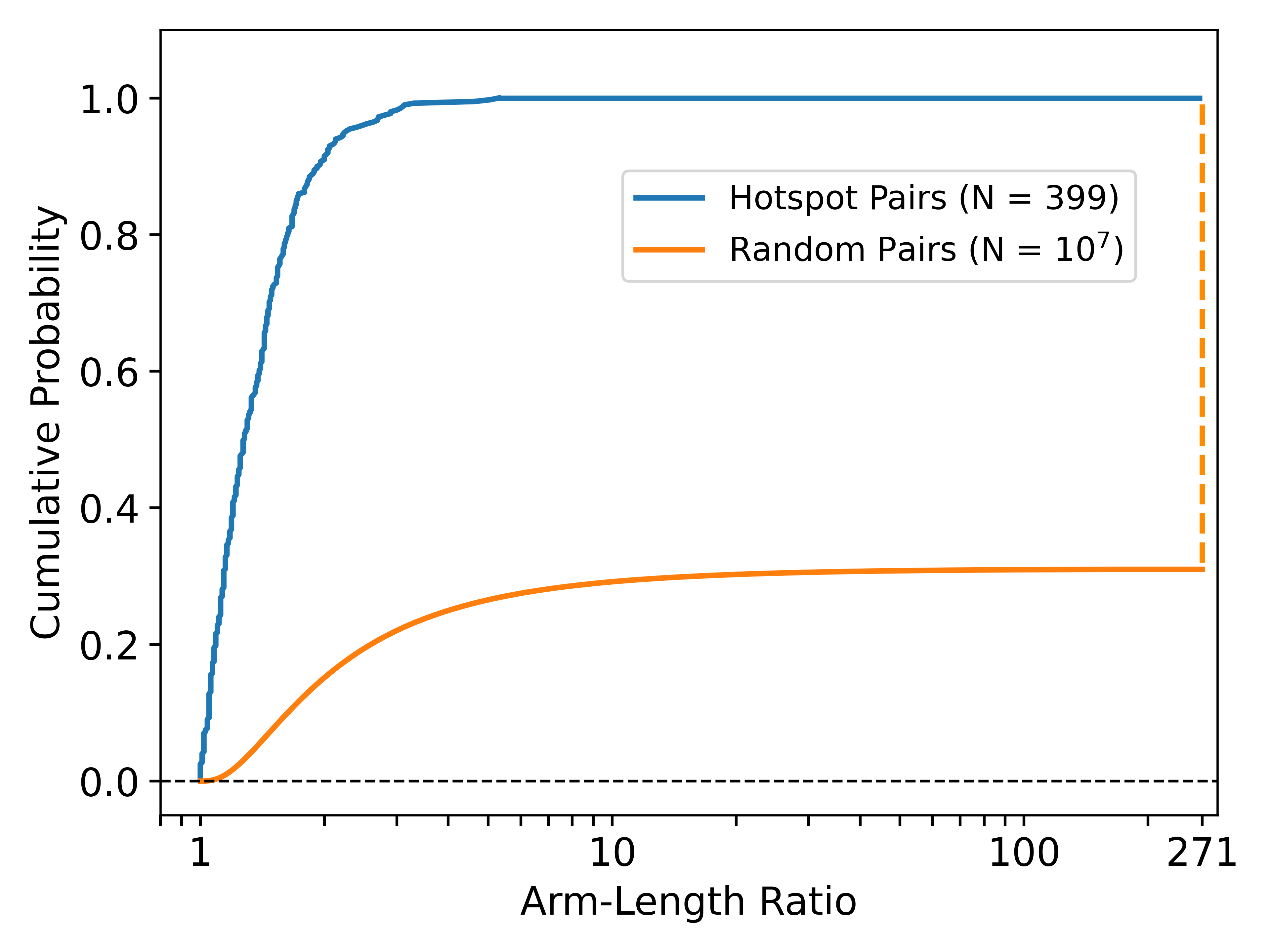}
    \caption{Cumulative probability distributions of the observed arm-length ratios for hotspot pairs in double radio galaxies (blue) and for randomly located component pairs generated by simulations (orange). The terminal jump in the simulated distribution comes from cases that did not have 3 components on each side, i.e. cases which do not contribute to the hotspot hypothesis.}
    \label{fig:probability}
\end{figure}

We next estimated the probabilities that the inner three pairs of compact components are either (i) hotspots from previous episodes of jet activity or (ii) randomly located plasma clouds along the jet path. We distinguish between the two cases using the arm-length ratio, defined as the ratio of the projected distances (from the radio core) of the farther hotspot to the nearer hotspot of a coeval pair.

We compared the observed arm-length ratios of the component pairs against the cumulative distributions constructed from (i) published data on 399 radio galaxies \citep{Macklin1981, McCarthy1991, Best1995, Ishwara-Chandra1999, Schoenmakers2000, Lara2004, Pirya2012, Andernach2025, Sethi2025}, and from (ii) $10^7$ Monte-Carlo simulations for 6 components randomly positioned within the two outermost hotspots relative to the radio core.

About $\sim$70.5\%\ of the Monte-Carlo trials did not yield equal number of components on the two sides of the radio core, i.e. they yielded 4+2 or 5+1 or 6+0 components. Since the observed components occur in equal numbers on both sides of the radio core these 70.5\% trials were counted as failures for the knots-in-the-jet scenario. The cumulative distributions for the double radio galaxy sample and the Monte-Carlo trials are shown in Figure~\ref{fig:probability}. From these, we estimated that the probability that these are 3 inner pairs of hotspots is $\sim3\times10^{6}$ times more than that of the knots-in-the-jet model.

Therefore, the multi-epoch hotspot model is supported by both the difference in spectral ages and by the locations of the hotspots (arm-length ratio).

Hence, we conclude that J023721.13$-$010528.5 exhibits four distinct episodes of jet activity, i.e. a quadruple-double radio galaxy (QDRG). This discovery highlights the importance of identifying and studying larger samples of highly recurrent radio galaxies. Expanding the currently small sample of these rare systems is crucial, as they preserve a long-term record of AGN activity and provide a unique opportunity to constrain the timescales and mechanisms governing jet launching, cessation, and reactivation (a.k.a. the duty cycle), as well as the interplay between AGN and their surroundings \citep{Morganti2021, Mahatma2023}. 

\begin{figure}
    \centering
    \includegraphics[width=\linewidth]{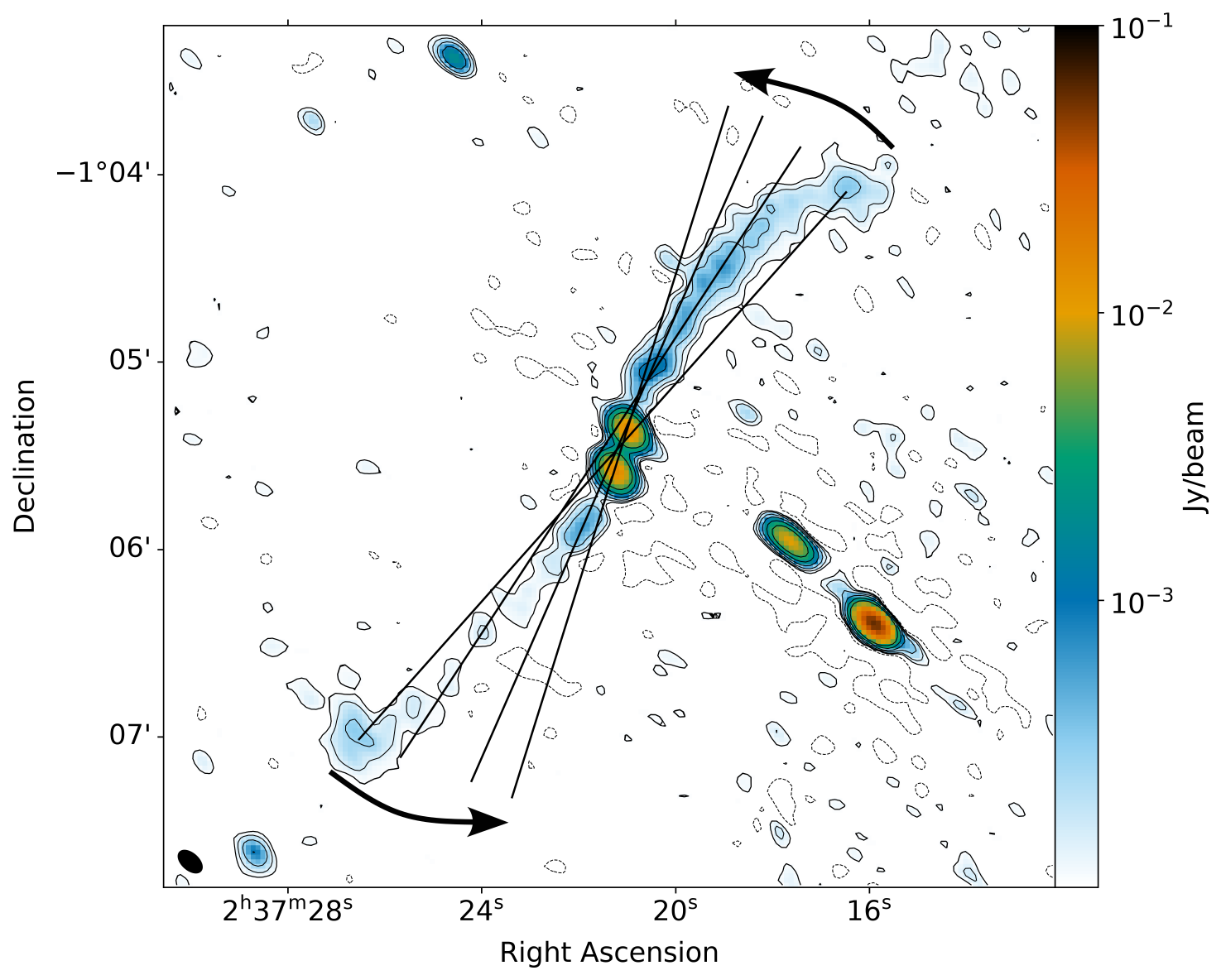}
    \caption{Change in orientation of the radio jet axes resulting in the $S$-shape. The jet axes are the lines joining the pairs of hotspots of the four episodes in the GMRT band-3 (400 MHz) image.}
    \label{fig:wobbling}
\end{figure}

\subsection{QDRG J023721.13$-$010528.5: Comparison with DDRGs, TDRGs and $S$-shaped Radio Galaxies}

QDRG J023721.13$-$010528.5 has a projected linear size of 1306~kpc. Its large size, together with the absence of broad emission lines in the optical spectrum, suggests that the jet axis lies close to the plane of the sky, or at least at an angle $\gtrsim 45^\circ$ to the line of sight \citep{Barthel1989}. Taking the innermost hotspot pair (N1--S1) as the reference, the position angles of the successive hotspot pairs rotate systematically in the counter-clockwise direction, with offsets of $-7^\circ$ for N2--S2, $-16^\circ$ for N3--S3, and $-24^\circ$ for N4--S4. This progressive change in orientation gives rise to the characteristic $S$-shaped radio morphology of QDRG J023721.13$-$010528.5 (see also Figure~\ref{fig:wobbling}).

\begin{figure}
    \centering
    \includegraphics[width=\linewidth]{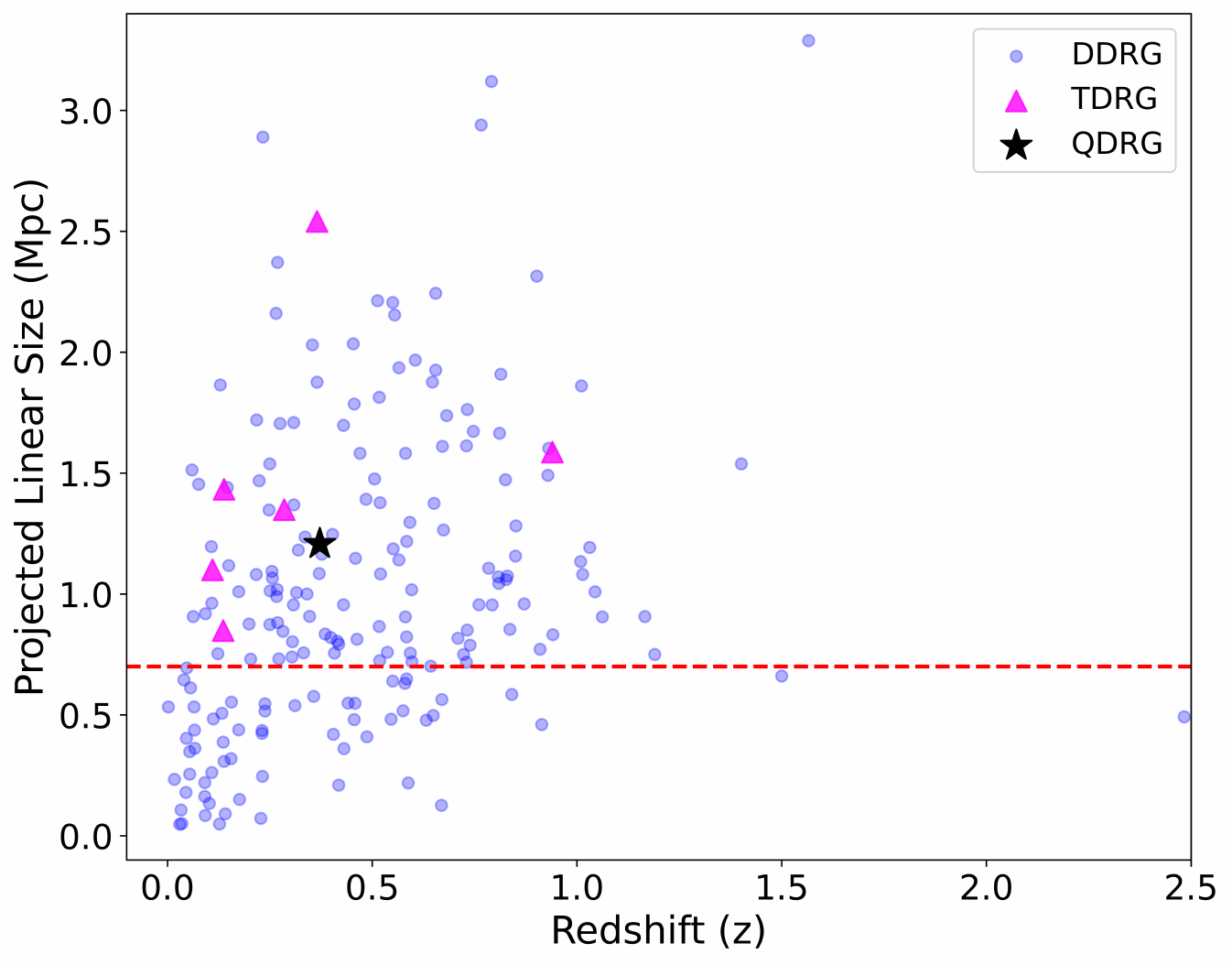}
    \caption{The plot displays the projected linear sizes (between the outermost hotspots) of different classes of episodic radio galaxies as a function of redshift: DDRGs (blue circles), TDRGs (magenta triangles), and the QDRG J023721.13$-$010528.5 (black star). The red dashed line marks the 700 kpc boundary for GRGs.
    }
    \label{fig:D_vs_z}
\end{figure}

In typical DDRGs showing episodic nuclear activity, the outer lobes are interpreted as relics of a previous episode of jet activity. Once the supply of fresh relativistic particles from the central engine ceases, these lobes evolve passively and their radio spectra progressively steepen. A subsequent reactivation of the AGN launches a new pair of jets, giving rise to the inner lobes. This interpretation is supported by spectral studies, which consistently show that the outer lobes have steeper radio spectra than the inner lobes, indicating that they contain older electron populations \citep{Schoenmaker2000,Konar2006}.

The north-eastern jet of QDRG J023721.13$-$010528.5 shows a clear trend towards increasing spectral age of a hotspot pair with its distance from the core. This is less so in the south-eastern jet, which is fainter and has larger measurement errors. Of the 192 DDRGs known, there are none in which the outer hotspot is more compact and younger, i.e. the active hotspot has never gone past the older hotspot. However, the inner hotspot is presumably still being energised by an active jet and we cannot predict how far it will progress. This pattern can only be definitively addressed in sources which have at least two non-active hotspots, i.e. radio sources with 3 or more epochs of jet activity. At present we have the spectral ages for just two such sources: the QDRG presented here, and J022248$-$060934 \citep{Rarivoarinoro2026}. Both sources show the pattern of increasing spectral age with distance. The simplest explanation for the pattern seen in multi-epoch radio sources taken as a whole is that either the jet power and/or the duration of the on-phase reduces with each successive episode. Whether or not this pattern is a feature of most, if not all, multi-epoch radio source will need a larger sample of TDRGs and QDRGs.

There appears to be no general pattern in the variation of flux density (and hence spectral luminosity) of hotspots with separation from the radio core across DDRGs and TDRGs. For example, in four of the six known giant TDRGs, the flux density increases toward the outer hotspots, whereas the remaining two systems, including QDRG J023721.13$-$010528.5, do not follow this trend. Symmetry properties also show significant diversity among TDRGs; for example, J0929$+$4146 becomes increasingly asymmetric toward the outer lobes \citep{Brocksopp2007}, while in other sources no clear trend is observed, and in some cases the direction of asymmetry even changes between successive episodes. In contrast, our QDRG shows an overall $S$-shaped symmetry across all four episodes.

\begin{figure}
    \centering
    \includegraphics[width=\linewidth]{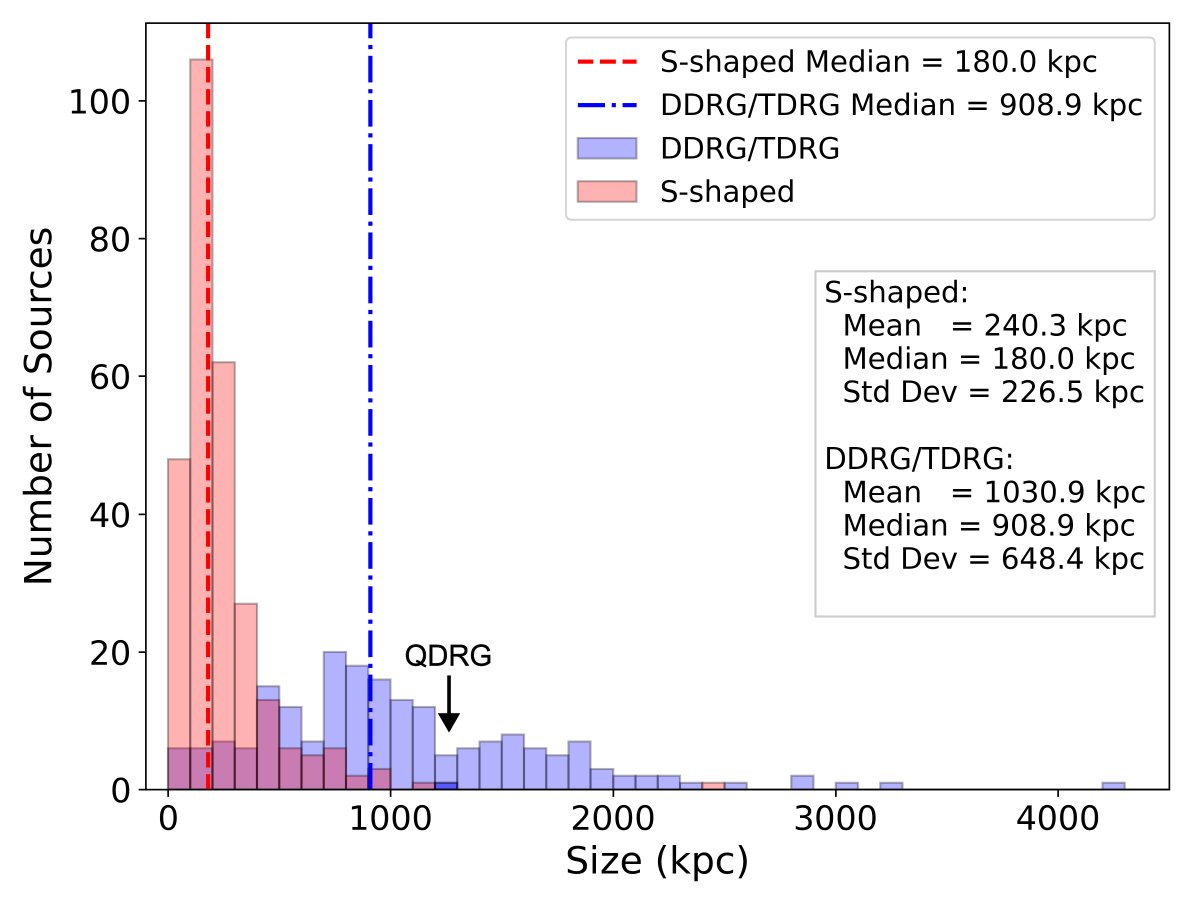}
    \caption{Figure showing the distribution of projected linear sizes (between the outermost hotspots) for different classes of episodic radio galaxies: The QDRG J023721.13$-$010528.5 is indicated by an arrow, $S$-shaped radio galaxies are in red, and DDRGs and TDRGs are shown in blue.
    The red dashed and blue dash-dotted lines indicate the median sizes for S-shaped radio galaxies, and of combined DDRGs and TDRGs, respectively. 
    The inset box summarizes key statistics, including the mean, median, and standard deviation for $S$-shaped radio galaxies, and DDRGs and TDRGs combined.
    }
    \label{fig:hist_LLS}
\end{figure}

Figure~\ref{fig:D_vs_z} shows a scatter plot of the projected linear sizes between the outermost hotspots of different classes of multi-episodic radio galaxies as a function of redshift. We used the hotspot separation instead of the distance between the farthest contours since the former is relatively uninfluenced by resolution and frequency of observation. The red dashed line in the plot marks the 700 kpc boundary for giant radio galaxies. We see that QDRG J023721.13$-$010528.5, even while being a giant radio galaxy, lies well within the observed distribution in terms of size and redshift. Figure~\ref{fig:hist_LLS} presents a histogram of the projected linear sizes of 280 $S$-shaped radio galaxies \citep{Ulvestad1983, Condon1984, Baum1988, Kotanyi1990, Giroletti2005, Hong2008, Solovyov2014, Liu2019, Bera2020, Bruno2024, Sethi2024, Ubertosi2024, Gopalkrishna2025, Misra2025} and 198 multi-epoch radio sources (cited earlier in this paper). The median projected size is 180~kpc for $S$-shaped sources (red-dashed line), and 908.9~kpc (blue dashed line) for sources with multi-episodic jet activity. QDRG J023721.13$-$010528.5, at 1.2~Mpc, is the second largest $S$-shaped radio source known.

The strong association of multi-episodic radio galaxies with larger radio structures may be a selection effect, as it is easier to separate and analyse the hotspots from multiple epochs in larger sources. The 6 known TDRGs and the QDRG presented here are all giant radio sources. Figure~\ref{fig:D_vs_z} shows a correlation between projected linear size and redshift even in DDRGs with an absence of small sources at high redshifts (larger distances), presumably due to insufficient resolution to separate the hotspots from multiple epochs. This ability to separate multi-epoch hotspots may be further enhanced in $S$-shaped sources, where precessing jets distribute the multi-episode hotspots over two dimensions. Therefore, the larger $S$-shaped sources may be good targets for increasing the number of known TDRGs and QDRGs.

\section{Conclusions}
J023721.13$-$010528.5 is one of the rarest of the rare radio source: it is a GRG with an $S$-shaped morphology and the first radio galaxy known to exhibit four distinct episodes of radio jet activity, a QDRG. Multi-frequency observations using uGMRT, and data from MeerKAT and VLA yielded radiative ages of the hotspots, which provide strong evidence for four distinct episodes of activity. They also showed a correlation between the hotspot age and the separation between corresponding hotspot pairs,
which suggests that the central engine powering the radio source becomes weaker with age.  
Thus, either the jet power decreases, the duration of jet activity in each episode decreases, or both, with no reason to believe that four episodes is the upper limit for a double radio source. Giant radio galaxies with $S$-shaped morphologies may be the best targets for identifying and studying multi-episodic radio activity, helping to understand the processes responsible for the largest known structures in the Universe.

\section*{acknowledgments}

We thank the anonymous referee for their comments, which have helped improve the paper.
The research of P.V.K. is supported by the University Grant Commission.
D.V.L. acknowledge the support of the Department of Atomic Energy, Government of India, under the project title ``Next Generation Instrumentation for Radio Astronomy" and the
project UID number RTI4017.
We thank the staff of the GMRT, which is run by the National Center for Radio Astrophysics of the Tata Institute of Fundamental Research. We are also grateful to the VLASS radio surveys, as well as the DESI, SDSS, and NED databases.

\section*{Data Availability}
The GMRT data underlying this article are available via the GMRT online archive facility\footnote{\url{https://naps.ncra.tifr.res.in/goa/data/search}}.
The MGCLS data used in this article are available on the MeerKAT SARAO website\footnote{\url{https://archive-gw-1.kat.ac.za/public/repository/10.48479/7epd-w356/index.html}}. The Desi Legacy Survey images are obtained from the Legacy Survey Viewer. The VLASS survey images are obtained from NRAO\footnote{\url{https://archive-new.nrao.edu/vlass/}}.

\facilities{GMRT, MeerKAT, VLA}

\software{CASA \citep{2022PASP..134k4501C}}

\vspace{5mm}

\newpage
\bibliography{references}{}

@article{FanaroffRiley1974,
    author = {Fanaroff, B. L. and Riley, J. M.},
    title = "{The Morphology of Extragalactic Radio Sources of High and Low Luminosity}",
    journal = {Monthly Notices of the Royal Astronomical Society},
    volume = {167},
    number = {1},
    pages = {31P-36P},
    year = {1974},
    month = {04},
    issn = {0035-8711},
    doi = {10.1093/mnras/167.1.31P},
    url = {https://doi.org/10.1093/mnras/167.1.31P},
    eprint = {https://academic.oup.com/mnras/article-pdf/167/1/31P/8079923/mnras167-031P.pdf},
}

@article{Konar2006,
    author = {Konar, C. and Saikia, D. J. and Jamrozy, M. and Machalski, J.},
    title = "{Spectral ageing analysis of the double–double radio galaxy J1453+3308}",
    journal = {Monthly Notices of the Royal Astronomical Society},
    volume = {372},
    number = {2},
    pages = {693-702},
    year = {2006},
    month = {09},
    issn = {0035-8711},
    doi = {10.1111/j.1365-2966.2006.10874.x},
    url = {https://doi.org/10.1111/j.1365-2966.2006.10874.x},
    eprint = {https://academic.oup.com/mnras/article-pdf/372/2/693/2996958/mnras0372-0693.pdf},
}

@article{Kaiser2000,
    author = {Kaiser, Christian R. and Schoenmakers, Arno P. and Röttgering, Huub J. A.},
    title = "{Radio galaxies with a ‘double-double’ morphology—II. The evolution of double-double radio galaxies and implications for the alignment effect in FRII sources}",
    journal = {Monthly Notices of the Royal Astronomical Society},
    volume = {315},
    number = {2},
    pages = {381-394},
    year = {2000},
    month = {06},
    issn = {0035-8711},
    doi = {10.1046/j.1365-8711.2000.03431.x},
    url = {https://doi.org/10.1046/j.1365-8711.2000.03431.x},
    eprint = {https://academic.oup.com/mnras/article-pdf/315/2/381/3042485/315-2-381.pdf},
}

@article{Brocksopp2007,
    author = {Brocksopp, C. and Kaiser, C. R. and Schoenmakers, A. P. and De Bruyn, A. G.},
    title = "{Three episodes of jet activity in the Fanaroff–Riley type II radio galaxy B0925+420}",
    journal = {Monthly Notices of the Royal Astronomical Society},
    volume = {382},
    number = {3},
    pages = {1019-1028},
    year = {2007},
    month = {11},
    issn = {0035-8711},
    doi = {10.1111/j.1365-2966.2007.12483.x},
    url = {https://doi.org/10.1111/j.1365-2966.2007.12483.x},
    eprint = {https://academic.oup.com/mnras/article-pdf/382/3/1019/18421381/mnras0382-1019.pdf},
}

@article{Hota2011,
    author = {Hota, Ananda and Sirothia, S. K. and Ohyama, Youichi and Konar, C. and Kim, Suk and Rey, Soo-Chang and Saikia, D. J. and Croston, J. H. and Matsushita, Satoki},
    title = "{Discovery of a spiral-host episodic radio galaxy}",
    journal = {Monthly Notices of the Royal Astronomical Society: Letters},
    volume = {417},
    number = {1},
    pages = {L36-L40},
    year = {2011},
    month = {10},
    issn = {1745-3925},
    doi = {10.1111/j.1745-3933.2011.01115.x},
    url = {https://doi.org/10.1111/j.1745-3933.2011.01115.x},
    eprint = {https://academic.oup.com/mnrasl/article-pdf/417/1/L36/56936074/mnrasl\_417\_1\_l36.pdf},
}

@article{Singh_2016,
doi = {10.3847/0004-637X/826/2/132},
url = {https://dx.doi.org/10.3847/0004-637X/826/2/132},
year = {2016},
month = {jul},
publisher = {The American Astronomical Society},
volume = {826},
number = {2},
pages = {132},
author = {Veeresh Singh and C. H. Ishwara-Chandra and Preeti Kharb and Shweta Srivastava and P. Janardhan},
title = {J1216+0709: A RADIO GALAXY WITH THREE EPISODES OF AGN JET ACTIVITY},
journal = {The Astrophysical Journal}
}

@ARTICLE{Chavan2023,
       author = {{Chavan}, Kshitij and {Dabhade}, Pratik and {Saikia}, D.~J.},
        title = "{A giant radio galaxy with three cycles of episodic jet activity from LoTSS DR2}",
      journal = {\mnras},
         year = 2023,
        month = oct,
       volume = {525},
       number = {1},
        pages = {L87-L92},
          doi = {10.1093/mnrasl/slad100},
archivePrefix = {arXiv},
       eprint = {2307.08553},
 primaryClass = {astro-ph.GA},
       adsurl = {https://ui.adsabs.harvard.edu/abs/2023MNRAS.525L..87C}
}

@ARTICLE{Swarup1991,
       author = {{Swarup}, G. and {Ananthakrishnan}, S. and {Kapahi}, V.~K. and {Rao}, A.~P. and {Subrahmanya}, C.~R. and {Kulkarni}, V.~K.},
        title = "{The Giant Metre-Wave Radio Telescope}",
      journal = {Current Science},
         year = 1991,
        month = jan,
       volume = {60},
        pages = {95},
       adsurl = {https://ui.adsabs.harvard.edu/abs/1991CSci...60...95S}
}

@ARTICLE{Gupta2017,
       author = {{Gupta}, Y. and {Ajithkumar}, B. and {Kale}, H.~S. and {Nayak}, S. and {Sabhapathy}, S. and {Sureshkumar}, S. and {Swami}, R.~V. and {Chengalur}, J.~N. and {Ghosh}, S.~K. and {Ishwara-Chandra}, C.~H. and {Joshi}, B.~C. and {Kanekar}, N. and {Lal}, D.~V. and {Roy}, S.},
        title = "{The upgraded GMRT: opening new windows on the radio Universe}",
      journal = {Current Science},
         year = 2017,
        month = aug,
       volume = {113},
       number = {4},
        pages = {707-714},
          doi = {10.18520/cs/v113/i04/707-714},
       adsurl = {https://ui.adsabs.harvard.edu/abs/2017CSci..113..707G}
}

@article{Tremaine_2002,
doi = {10.1086/341002},
url = {https://dx.doi.org/10.1086/341002},
year = {2002},
month = {aug},
publisher = {},
volume = {574},
number = {2},
pages = {740},
author = {Scott Tremaine and Karl Gebhardt and Ralf Bender and Gary Bower and Alan Dressler and S. M. Faber and Alexei V. Filippenko and Richard Green and Carl Grillmair and Luis C. Ho and John Kormendy and Tod R. Lauer and John Magorrian and Jason Pinkney and Douglas Richstone},
title = {The Slope of the Black Hole Mass versus Velocity Dispersion Correlation},
journal = {The Astrophysical Journal}
}

@article{Schoenmaker2000,
author = {Schoenmakers, A. and de Bruyn, A. and Röttgering, H. and Laan, H.},
year = {2000},
month = {06},
pages = {395 - 406},
title = {Radio galaxies with a ‘double‐double’ morphology – III. The case of B 1834+620},
volume = {315},
journal = {Monthly Notices of the Royal Astronomical Society},
doi = {10.1046/j.1365-8711.2000.03432.x}
}

@ARTICLE{Miley1980,
       author = {{Miley}, G.},
        title = "{The structure of extended extragalactic radio sources}",
      journal = {\araa},
         year = 1980,
        month = jan,
       volume = {18},
        pages = {165-218},
          doi = {10.1146/annurev.aa.18.090180.001121},
       adsurl = {https://ui.adsabs.harvard.edu/abs/1980ARA&A..18..165M}
}

@ARTICLE{Lal2013,
       author = {{Lal}, Dharam V. and {Kraft}, Ralph P. and {Randall}, Scott W. and {Forman}, William R. and {Nulsen}, Paul E.~J. and {Roediger}, Elke and {ZuHone}, John A. and {Hardcastle}, Martin J. and {Jones}, Christine and {Croston}, Judith H.},
        title = "{Gas Sloshing and Radio Galaxy Dynamics in the Core of the 3C 449 Group}",
      journal = {\apj},
         year = 2013,
        month = feb,
       volume = {764},
       number = {1},
          eid = {83},
        pages = {83},
          doi = {10.1088/0004-637X/764/1/83},
archivePrefix = {arXiv},
       eprint = {1210.7563},
 primaryClass = {astro-ph.CO},
       adsurl = {https://ui.adsabs.harvard.edu/abs/2013ApJ...764...83L}
}

@article{Lacy_2020,
   title={The Karl G. Jansky Very Large Array Sky Survey (VLASS). Science Case and Survey Design},
   volume={132},
   ISSN={1538-3873},
   url={http://dx.doi.org/10.1088/1538-3873/ab63eb},
   DOI={10.1088/1538-3873/ab63eb},
   number={1009},
   journal={Publications of the Astronomical Society of the Pacific},
   publisher={IOP Publishing},
   author={Lacy, M. and Baum, S. A. and Chandler, C. J. and Chatterjee, S. and Clarke, T. E. and Deustua, S. and English, J. and Farnes, J. and Gaensler, B. M. and Gugliucci, N. and Hallinan, G. and Kent, B. R. and Kimball, A. and Law, C. J. and Lazio, T. J. W. and Marvil, J. and Mao, S. A. and Medlin, D. and Mooley, K. and Murphy, E. J. and Myers, S. and Osten, R. and Richards, G. T. and Rosolowsky, E. and Rudnick, L. and Schinzel, F. and Sivakoff, G. R. and Sjouwerman, L. O. and Taylor, R. and White, R. L. and Wrobel, J. and Andernach, H. and Beasley, A. J. and Berger, E. and Bhatnager, S. and Birkinshaw, M. and Bower, G. C. and Brandt, W. N. and Brown, S. and Burke-Spolaor, S. and Butler, B. J. and Comerford, J. and Demorest, P. B. and Fu, H. and Giacintucci, S. and Golap, K. and Güth, T. and Hales, C. A. and Hiriart, R. and Hodge, J. and Horesh, A. and Ivezić, Ž. and Jarvis, M. J. and Kamble, A. and Kassim, N. and Liu, X. and Loinard, L. and Lyons, D. K. and Masters, J. and Mezcua, M. and Moellenbrock, G. A. and Mroczkowski, T. and Nyland, K. and O’Dea, C. P. and O’Sullivan, S. P. and Peters, W. M. and Radford, K. and Rao, U. and Robnett, J. and Salcido, J. and Shen, Y. and Sobotka, A. and Witz, S. and Vaccari, M. and Weeren, R. J. van and Vargas, A. and Williams, P. K. G. and Yoon, I.},
   year={2020},
   month=jan, pages={035001} }

@software{SPAM2014,
       author = {{Intema}, Huib T.},
        title = "{SPAM: Source Peeling and Atmospheric Modeling}",
 howpublished = {Astrophysics Source Code Library, record ascl:1408.006},
         year = 2014,
        month = aug,
          eid = {ascl:1408.006},
       adsurl = {https://ui.adsabs.harvard.edu/abs/2014ascl.soft08006I}
}

@ARTICLE{Kardashev1962,
       author = {{Kardashev}, N.~S.},
        title = "{Nonstationarity of Spectra of Young Sources of Nonthermal Radio Emission}",
      journal = {\sovast},
         year = 1962,
        month = dec,
       volume = {6},
        pages = {317},
       adsurl = {https://ui.adsabs.harvard.edu/abs/1962SvA.....6..317K}
}

@ARTICLE{Dabhade2020b,
       author = {{Dabhade}, P. and {Mahato}, M. and {Bagchi}, J. and {Saikia}, D.~J. and {Combes}, F. and {Sankhyayan}, S. and {R{\"o}ttgering}, H.~J.~A. and {Ho}, L.~C. and {Gaikwad}, M. and {Raychaudhury}, S. and {Vaidya}, B. and {Guiderdoni}, B.},
        title = "{Search and analysis of giant radio galaxies with associated nuclei (SAGAN). I. New sample and multi-wavelength studies}",
      journal = {\aap},
         year = 2020,
        month = oct,
       volume = {642},
          eid = {A153},
        pages = {A153},
          doi = {10.1051/0004-6361/202038344},
archivePrefix = {arXiv},
       eprint = {2005.03708},
 primaryClass = {astro-ph.GA},
       adsurl = {https://ui.adsabs.harvard.edu/abs/2020A&A...642A.153D}
}

@ARTICLE{Yuan2016,
       author = {{Yuan}, Z.~S. and {Han}, J.~L. and {Wen}, Z.~L.},
        title = "{Radio luminosity function of brightest cluster galaxies}",
      journal = {\mnras},
         year = 2016,
        month = aug,
       volume = {460},
       number = {4},
        pages = {3669-3678},
          doi = {10.1093/mnras/stw1125},
archivePrefix = {arXiv},
       eprint = {1605.03387},
 primaryClass = {astro-ph.GA},
       adsurl = {https://ui.adsabs.harvard.edu/abs/2016MNRAS.460.3669Y}
}

@ARTICLE{Knowles2022,
       author = {{Knowles}, K. and {Cotton}, W.~D. and {Rudnick}, L. and {Camilo}, F. and {Goedhart}, S. and {Deane}, R. and {Ramatsoku}, M. and {Bietenholz}, M.~F. and {Br{\"u}ggen}, M. and {Button}, C. and {Chen}, H. and {Chibueze}, J.~O. and {Clarke}, T.~E. and {de Gasperin}, F. and {Ianjamasimanana}, R. and {J{\'o}zsa}, G.~I.~G. and {Hilton}, M. and {Kesebonye}, K.~C. and {Kolokythas}, K. and {Kraan-Korteweg}, R.~C. and {Lawrie}, G. and {Lochner}, M. and {Loubser}, S.~I. and {Marchegiani}, P. and {Mhlahlo}, N. and {Moodley}, K. and {Murphy}, E. and {Namumba}, B. and {Oozeer}, N. and {Parekh}, V. and {Pillay}, D.~S. and {Passmoor}, S.~S. and {Ramaila}, A.~J.~T. and {Ranchod}, S. and {Retana-Montenegro}, E. and {Sebokolodi}, L. and {Sikhosana}, S.~P. and {Smirnov}, O. and {Thorat}, K. and {Venturi}, T. and {Abbott}, T.~D. and {Adam}, R.~M. and {Adams}, G. and {Aldera}, M.~A. and {Bauermeister}, E.~F. and {Bennett}, T.~G.~H. and {Bode}, W.~A. and {Botha}, D.~H. and {Botha}, A.~G. and {Brederode}, L.~R.~S. and {Buchner}, S. and {Burger}, J.~P. and {Cheetham}, T. and {de Villiers}, D.~I.~L. and {Dikgale-Mahlakoana}, M.~A. and {du Toit}, L.~J. and {Esterhuyse}, S.~W.~P. and {Fadana}, G. and {Fanaroff}, B.~L. and {Fataar}, S. and {Foley}, A.~R. and {Fourie}, D.~J. and {Frank}, B.~S. and {Gamatham}, R.~R.~G. and {Gatsi}, T.~G. and {Geyer}, M. and {Gouws}, M. and {Gumede}, S.~C. and {Heywood}, I. and {Hlakola}, M.~J. and {Hokwana}, A. and {Hoosen}, S.~W. and {Horn}, D.~M. and {Horrell}, J.~M.~G. and {Hugo}, B.~V. and {Isaacson}, A.~R. and {Jonas}, J.~L. and {Jordaan}, J.~D.~B. and {Joubert}, A.~F. and {Julie}, R.~P.~M. and {Kapp}, F.~B. and {Kasper}, V.~A. and {Kenyon}, J.~S. and {Kotz{\'e}}, P.~P.~A. and {Kotze}, A.~G. and {Kriek}, N. and {Kriel}, H. and {Krishnan}, V.~K. and {Kusel}, T.~W. and {Legodi}, L.~S. and {Lehmensiek}, R. and {Liebenberg}, D. and {Lord}, R.~T. and {Lunsky}, B.~M. and {Madisa}, K. and {Magnus}, L.~G. and {Main}, J.~P.~L. and {Makhaba}, A. and {Makhathini}, S. and {Malan}, J.~A. and {Manley}, J.~R. and {Marais}, S.~J. and {Maree}, M.~D.~J. and {Martens}, A. and {Mauch}, T. and {McAlpine}, K. and {Merry}, B.~C. and {Millenaar}, R.~P. and {Mokone}, O.~J. and {Monama}, T.~E. and {Mphego}, M.~C. and {New}, W.~S. and {Ngcebetsha}, B. and {Ngoasheng}, K.~J. and {Ockards}, M.~T. and {Otto}, A.~J. and {Patel}, A.~A. and {Peens-Hough}, A. and {Perkins}, S.~J. and {Ramanujam}, N.~M. and {Ramudzuli}, Z.~R. and {Ratcliffe}, S.~M. and {Renil}, R. and {Robyntjies}, A. and {Rust}, A.~N. and {Salie}, S. and {Sambu}, N. and {Schollar}, C.~T.~G. and {Schwardt}, L.~C. and {Schwartz}, R.~L. and {Serylak}, M. and {Siebrits}, R. and {Sirothia}, S.~K. and {Slabber}, M. and {Sofeya}, L. and {Taljaard}, B. and {Tasse}, C. and {Tiplady}, A.~J. and {Toruvanda}, O. and {Twum}, S.~N. and {van Balla}, T.~J. and {van der Byl}, A. and {van der Merwe}, C. and {van Dyk}, C.~L. and {Van Tonder}, V. and {Van Wyk}, R. and {Venter}, A.~J. and {Venter}, M. and {Welz}, M.~G. and {Williams}, L.~P. and {Xaia}, B.},
        title = "{The MeerKAT Galaxy Cluster Legacy Survey. I. Survey Overview and Highlights}",
      journal = {\aap},
         year = 2022,
        month = jan,
       volume = {657},
          eid = {A56},
        pages = {A56},
          doi = {10.1051/0004-6361/202141488},
archivePrefix = {arXiv},
       eprint = {2111.05673},
 primaryClass = {astro-ph.GA},
       adsurl = {https://ui.adsabs.harvard.edu/abs/2022A&A...657A..56K}
}

@article{Kormendy2013,
   author = "Kormendy, John and Ho, Luis C.",
   title = "Coevolution (Or Not) of Supermassive Black Holes and Host Galaxies", 
   journal= "Annual Review of Astronomy and Astrophysics",
   year = "2013",
   volume = "51",
   number = "Volume 51, 2013",
   pages = "511-653",
   doi = "https://doi.org/10.1146/annurev-astro-082708-101811",
   url = "https://www.annualreviews.org/content/journals/10.1146/annurev-astro-082708-101811",
   publisher = "Annual Reviews",
   issn = "1545-4282",
   type = "Journal Article",
  }

@ARTICLE{Barthel1989,
       author = {{Barthel}, Peter D.},
        title = "{Is Every Quasar Beamed?}",
      journal = {\apj},
         year = 1989,
        month = jan,
       volume = {336},
        pages = {606},
          doi = {10.1086/167038},
       adsurl = {https://ui.adsabs.harvard.edu/abs/1989ApJ...336..606B}
}

@ARTICLE{Mahatma2019,
       author = {{Mahatma}, V.~H. and {Hardcastle}, M.~J. and {Williams}, W.~L. and {Best}, P.~N. and {Croston}, J.~H. and {Duncan}, K. and {Mingo}, B. and {Morganti}, R. and {Brienza}, M. and {Cochrane}, R.~K. and {G{\"u}rkan}, G. and {Harwood}, J.~J. and {Jarvis}, M.~J. and {Jamrozy}, M. and {Jurlin}, N. and {Morabito}, L.~K. and {R{\"o}ttgering}, H.~J.~A. and {Sabater}, J. and {Shimwell}, T.~W. and {Smith}, D.~J.~B. and {Shulevski}, A. and {Tasse}, C.},
        title = "{LoTSS DR1: Double-double radio galaxies in the HETDEX field}",
      journal = {\aap},
         year = 2019,
        month = feb,
       volume = {622},
          eid = {A13},
        pages = {A13},
          doi = {10.1051/0004-6361/201833973},
archivePrefix = {arXiv},
       eprint = {1811.08194},
 primaryClass = {astro-ph.GA},
       adsurl = {https://ui.adsabs.harvard.edu/abs/2019A&A...622A..13M}
}

@ARTICLE{Kuzmicz2017,
       author = {{Ku{\'z}micz}, A. and {Jamrozy}, M. and {Kozie{\l}-Wierzbowska}, D. and {We{\.z}gowiec}, M.},
        title = "{Optical and radio properties of extragalactic radio sources with recurrent jet activity}",
      journal = {\mnras},
         year = 2017,
        month = nov,
       volume = {471},
       number = {4},
        pages = {3806-3826},
          doi = {10.1093/mnras/stx1830},
archivePrefix = {arXiv},
       eprint = {1709.01802},
 primaryClass = {astro-ph.GA},
       adsurl = {https://ui.adsabs.harvard.edu/abs/2017MNRAS.471.3806K}
}

@ARTICLE{Nandi2012,
       author = {{Nandi}, S. and {Saikia}, D.~J.},
        title = "{Double-double radio galaxies from the FIRST survey}",
      journal = {Bulletin of the Astronomical Society of India},
         year = 2012,
        month = jun,
       volume = {40},
       number = {2},
        pages = {121-137},
          doi = {10.48550/arXiv.1208.1941},
archivePrefix = {arXiv},
       eprint = {1208.1941},
 primaryClass = {astro-ph.CO},
       adsurl = {https://ui.adsabs.harvard.edu/abs/2012BASI...40..121N}
}

@ARTICLE{Beck2005,
       author = {{Beck}, R. and {Krause}, M.},
        title = "{Revised equipartition and minimum energy formula for magnetic field strength estimates from radio synchrotron observations}",
      journal = {Astronomische Nachrichten},
         year = 2005,
        month = jul,
       volume = {326},
       number = {6},
        pages = {414-427},
          doi = {10.1002/asna.200510366},
archivePrefix = {arXiv},
       eprint = {astro-ph/0507367},
 primaryClass = {astro-ph},
       adsurl = {https://ui.adsabs.harvard.edu/abs/2005AN....326..414B}
}

@article{Kaiser1997,
    author = {Kaiser, Christian R. and Alexander, Paul},
    title = {A self-similar model for extragalactic radio sources},
    journal = {Monthly Notices of the Royal Astronomical Society},
    volume = {286},
    number = {1},
    pages = {215-222},
    year = {1997},
    month = {03},
    issn = {0035-8711},
    doi = {10.1093/mnras/286.1.215},
    url = {https://doi.org/10.1093/mnras/286.1.215},
    eprint = {https://academic.oup.com/mnras/article-pdf/286/1/215/5553393/286-1-215.pdf},
}

@article{Wen2015,
doi = {10.1088/0004-637X/807/2/178},
url = {https://dx.doi.org/10.1088/0004-637X/807/2/178},
year = {2015},
month = {jul},
publisher = {The American Astronomical Society},
volume = {807},
number = {2},
pages = {178},
author = {Wen, Z. L. and Han, J. L.},
title = {CALIBRATION OF THE OPTICAL MASS PROXY FOR CLUSTERS OF GALAXIES AND AN UPDATE OF THE WHL12 CLUSTER CATALOG},
journal = {The Astrophysical Journal}
}

@ARTICLE{Dabhade2025,
       author = {{Dabhade}, P. and {Chavan}, K. and {Saikia}, D.~J. and {Oei}, M.~S.~S.~L. and {R{\"o}ttgering}, H.~J.~A.},
        title = "{Search and analysis of giant radio galaxies with associated nuclei (SAGAN): V. Study of giant double-double radio galaxies from LoTSS DR2}",
      journal = {\aap},
         year = 2025,
        month = apr,
       volume = {696},
          eid = {A97},
        pages = {A97},
          doi = {10.1051/0004-6361/202451918},
archivePrefix = {arXiv},
       eprint = {2408.13607},
 primaryClass = {astro-ph.GA},
       adsurl = {https://ui.adsabs.harvard.edu/abs/2025A&A...696A..97D}
}

@article{Liu2019,
author = {Liu, Yu-Xing and Xu, Hai-Guang and Zheng, Dong-Chao and Li, Wei-Tian and Zhu, Zheng-Hao and Ma, Zhixian and Lian, Xiao-Li},
year = {2019},
month = {09},
pages = {127},
title = {The environment of C- and S-shaped radio galaxies},
volume = {19},
journal = {Research in Astronomy and Astrophysics},
doi = {10.1088/1674-4527/19/9/127}
}

@article{Bera2020,
doi = {10.3847/1538-4365/abb367},
url = {https://dx.doi.org/10.3847/1538-4365/abb367},
year = {2020},
month = {nov},
publisher = {The American Astronomical Society},
volume = {251},
number = {1},
pages = {9},
author = {Bera, Soumen and Pal, Sabyasachi and Sasmal, Tapan K. and Mondal, Soumen},
title = {FIRST Winged Radio Galaxies with X and Z Symmetry},
journal = {The Astrophysical Journal Supplement Series}
}

@misc{Gopalkrishna2025,
      title={A sample of 25 radio galaxies with highly unusual radio morphologies, selected from the LoTSS-DR2 survey at 144 MHz}, 
      author={Gopal-Krishna and Dusmanta Patra and Ravi Joshi},
      year={2025},
      eprint={2403.11290},
      archivePrefix={arXiv},
      primaryClass={astro-ph.GA},
      url={https://arxiv.org/abs/2403.11290}, 
}

@article{Hong2008,
doi = {10.1088/1009-9271/8/2/05},
url = {https://dx.doi.org/10.1088/1009-9271/8/2/05},
year = {2008},
month = {apr},
publisher = {},
volume = {8},
number = {2},
pages = {179},
author = {Xiao-Yu Hong and Chuan-Hao Sun and Jun-Hui Zhao and Dong-Rong Jiang and Zhi-Qiang Shen and Tao An and Wei-Hua Wang and Jun Yang},
title = {Bending of Jets in the QSO NRAO 530},
journal = {Chinese Journal of Astronomy and Astrophysics }
}

@article{Misra2025,
    author = {Misra, Arpita and Jamrozy, Marek and Weżgowiec, Marek and Kozieł-Wierzbowska, Dorota},
    title = {Multiwavelength investigations of PKS 2300–18: S-shaped radio quasar with precessing jets and double-peaked broad emission-line spectrum},
    journal = {Monthly Notices of the Royal Astronomical Society},
    volume = {536},
    number = {3},
    pages = {2025-2045},
    year = {2024},
    month = {11},
    issn = {0035-8711},
    doi = {10.1093/mnras/stae2639},
    url = {https://doi.org/10.1093/mnras/stae2639},
    eprint = {https://academic.oup.com/mnras/article-pdf/536/3/2025/60828611/stae2639.pdf},
}

@ARTICLE{2022PASP..134k4501C,
       author = {{CASA Team} and {Bean}, Ben and {Bhatnagar}, Sanjay and {Castro}, Sandra and {Donovan Meyer}, Jennifer and {Emonts}, Bjorn and {Garcia}, Enrique and {Garwood}, Robert and {Golap}, Kumar and {Gonzalez Villalba}, Justo and {Harris}, Pamela and {Hayashi}, Yohei and {Hoskins}, Josh and {Hsieh}, Mingyu and {Jagannathan}, Preshanth and {Kawasaki}, Wataru and {Keimpema}, Aard and {Kettenis}, Mark and {Lopez}, Jorge and {Marvil}, Joshua and {Masters}, Joseph and {McNichols}, Andrew and {Mehringer}, David and {Miel}, Renaud and {Moellenbrock}, George and {Montesino}, Federico and {Nakazato}, Takeshi and {Ott}, Juergen and {Petry}, Dirk and {Pokorny}, Martin and {Raba}, Ryan and {Rau}, Urvashi and {Schiebel}, Darrell and {Schweighart}, Neal and {Sekhar}, Srikrishna and {Shimada}, Kazuhiko and {Small}, Des and {Steeb}, Jan-Willem and {Sugimoto}, Kanako and {Suoranta}, Ville and {Tsutsumi}, Takahiro and {van Bemmel}, Ilse M. and {Verkouter}, Marjolein and {Wells}, Akeem and {Xiong}, Wei and {Szomoru}, Arpad and {Griffith}, Morgan and {Glendenning}, Brian and {Kern}, Jeff},
        title = "{CASA, the Common Astronomy Software Applications for Radio Astronomy}",
      journal = {\pasp},
         year = 2022,
        month = nov,
       volume = {134},
       number = {1041},
          eid = {114501},
        pages = {114501},
          doi = {10.1088/1538-3873/ac9642},
archivePrefix = {arXiv},
       eprint = {2210.02276},
 primaryClass = {astro-ph.IM},
       adsurl = {https://ui.adsabs.harvard.edu/abs/2022PASP..134k4501C}
}

@ARTICLE{Baum1988,
       author = {{Baum}, Stefi Alison and {Heckman}, Timothy M. and {Bridle}, Alan and {van Breugel}, Wil J.~M. and {Miley}, George K.},
        title = "{Extended Optical Line Emitting Gas in Radio Galaxies: Broad-Band Optical, Narrow-Band Optical, and Radio Imaging of a Representative Sample}",
      journal = {\apjs},
         year = 1988,
        month = dec,
       volume = {68},
        pages = {643},
          doi = {10.1086/191301},
       adsurl = {https://ui.adsabs.harvard.edu/abs/1988ApJS...68..643B}
}

@ARTICLE{Bruno2024,
       author = {{Bruno}, L. and {Brienza}, M. and {Zanichelli}, A. and {Gitti}, M. and {Ubertosi}, F. and {Rajpurohit}, K. and {Venturi}, T. and {Dallacasa}, D.},
        title = "{From 100 MHz to 10 GHz: Unveiling the spectral evolution of the X-shaped radio galaxy in Abell 3670}",
      journal = {\aap},
         year = 2024,
        month = oct,
       volume = {690},
          eid = {A160},
        pages = {A160},
          doi = {10.1051/0004-6361/202451682},
archivePrefix = {arXiv},
       eprint = {2408.11377},
 primaryClass = {astro-ph.GA},
       adsurl = {https://ui.adsabs.harvard.edu/abs/2024A&A...690A.160B}
}

@ARTICLE{Solovyov2014,
       author = {{Solovyov}, D.~I. and {Verkhodanov}, O.~V.},
        title = "{Radio galaxies with signatures of merging from the list of giant radio galaxy candidates based on NVSS data.}",
      journal = {Astronomy Letters},
         year = 2014,
        month = oct,
       volume = {40},
        pages = {606-614},
          doi = {10.1134/S1063773714090023},
       adsurl = {https://ui.adsabs.harvard.edu/abs/2014AstL...40..606S}
}

@ARTICLE{Ulvestad1983,
       author = {{Ulvestad}, J.~S. and {Wilson}, A.~S.},
        title = "{The nuclear radio source of the X-ray galaxy NGC 2110.}",
      journal = {\apjl},
         year = 1983,
        month = jan,
       volume = {264},
        pages = {L7-L11},
          doi = {10.1086/183935},
       adsurl = {https://ui.adsabs.harvard.edu/abs/1983ApJ...264L...7U}
}

@article{Sethi2024,
doi = {10.3847/1538-4357/ad500e},
url = {https://dx.doi.org/10.3847/1538-4357/ad500e},
year = {2024},
month = {jul},
publisher = {The American Astronomical Society},
volume = {969},
number = {2},
pages = {156},
author = {Sethi, Sagar and Kuźmicz, Agnieszka and Jamrozy, Marek and Slavcheva-Mihova, Lyuba},
title = {Discovery of a 100 kpc Narrow Curved Twin Jet in the S-shaped Giant Radio Galaxy J0644+1043},
journal = {The Astrophysical Journal}
}

@article{Ubertosi2024,
author = {{Ubertosi, F.} and {Giroletti, M.} and {Gitti, M.} and {Biava, N.} and {DeRubeis, E.} and {Bonafede, A.} and {Feretti, L.} and {Bondi, M.} and {Bruno, L.} and {Liuzzo, E.} and {Ignesti, A.} and {Brunetti, G.}},
title = {A JVLA, LOFAR, e-Merlin, VLBA, and EVN study of RBS 797: can binary supermassive black holes explain the outburst history of the central radio galaxy?},
DOI= "10.1051/0004-6361/202349011",
url= "https://doi.org/10.1051/0004-6361/202349011",
journal = {A\&A},
year = 2024,
volume = 688,
pages = "A86",
}

@ARTICLE{Kotanyi1990,
    author = {{Kotanyi}, C.},
    title = "{NGC 3309: an S-shaped radio galaxy in a nearby cluster.}",
    journal = {\rmxaa},
    year = 1990,
    month = nov,
    volume = {21},
    pages = {173},
    adsurl = {https://ui.adsabs.harvard.edu/abs/1990RMxAA..21..173K}
}

@article{Giroletti2005,
   title={The Two‐sided Parsec‐Scale Structure of the Low‐Luminosity Active Galactic Nucleus in NGC 4278},
   volume={622},
   ISSN={1538-4357},
   url={http://dx.doi.org/10.1086/427898},
   DOI={10.1086/427898},
   number={1},
   journal={The Astrophysical Journal},
   publisher={American Astronomical Society},
   author={Giroletti, M. and Taylor, G. B. and Giovannini, G.},
   year={2005},
   month=mar, pages={178–186} }

@ARTICLE{Condon1984,
       author = {{Condon}, J.~J. and {Mitchell}, K.~J.},
        title = "{4C 29.47 : Quasi-periodic outbursts recorded by precessing jets ?}",
      journal = {\apj},
         year = 1984,
        month = jan,
       volume = {276},
        pages = {472-475},
          doi = {10.1086/161634},
       adsurl = {https://ui.adsabs.harvard.edu/abs/1984ApJ...276..472C}
}

@software{BRATS,
       author = {{Harwood}, Jeremy J. and {Hardcastle}, Martin J. and {Croston}, Judith H. and {Goodger}, Joanna L.},
        title = "{BRATS: Broadband Radio Astronomy ToolS}",
 howpublished = {Astrophysics Source Code Library, record ascl:1806.025},
         year = 2018,
        month = jun,
          eid = {ascl:1806.025},
archivePrefix = {ascl},
       eprint = {1806.025},
       adsurl = {https://ui.adsabs.harvard.edu/abs/2018ascl.soft06025H}
}

@article{Morganti2014,
   title={Radio jets clearing the way through galaxies: the view from <scp>Hi</scp> and molecular gas},
   volume={10},
   ISSN={1743-9221},
   url={http://dx.doi.org/10.1017/S1743921315002331},
   DOI={10.1017/s1743921315002331},
   number={S313},
   journal={Proceedings of the International Astronomical Union},
   publisher={Cambridge University Press (CUP)},
   author={Morganti, Raffaella},
   year={2014},
   month=sep, pages={283–288} }

@ARTICLE{Fanaroff2021,
       author = {{Fanaroff}, Bernie and {Lal}, Dharam V. and {Venturi}, Tiziana and {Smirnov}, Oleg M. and {Bondi}, Marco and {Thorat}, Kshitij and {Bester}, Landman H. and {J{\'o}zsa}, Gyula I.~G. and {Kleiner}, Dane and {Loi}, Francesca and {Makhathini}, Sphesihle and {White}, Sarah V.},
        title = "{A new look at old friends - I. Imaging classical radio galaxies with uGMRT and MeerKAT}",
      journal = {\mnras},
         year = 2021,
        month = aug,
       volume = {505},
       number = {4},
        pages = {6003-6016},
          doi = {10.1093/mnras/stab1540},
archivePrefix = {arXiv},
       eprint = {2105.11695},
 primaryClass = {astro-ph.GA},
       adsurl = {https://ui.adsabs.harvard.edu/abs/2021MNRAS.505.6003F}
}

@misc{Rarivoarinoro2026,
      title={MIGHTEE: Discovery of a triple-double radio galaxy}, 
      author={Tombo F. Rarivoarinoro and Zara Randriamanakoto and Russ Taylor and Marisa Brienza and Fabio Luchsinger and Sushant Dutta and Catherine Hale and Jacinta Delhaize and Ndivhuwo Netshiavha and Solohery Randriamampandry and Mattia Vaccari},
      year={2026},
      eprint={2602.19729},
      archivePrefix={arXiv},
      primaryClass={astro-ph.GA},
      url={https://arxiv.org/abs/2602.19729}, 
}

@article{Norris2025, 
      title={EMU and the DRAGNs I: A catalogue of DRAGNs}, volume={42}, DOI={10.1017/pasa.2025.10076}, journal={Publications of the Astronomical Society of Australia}, author={Norris, Ray P. and Yew, Miranda and Crawford, Evan J. and Gupta, Nikhel and Rudnick, Lawrence and Andernach, Heinz and Filipović, Miroslav D. and Gordon, Yjan and Hopkins, Andrew and Park, Laurence and et al.}, year={2025}, pages={e124}
}

@ARTICLE{JP,
       author = {{Jaffe}, W.~J. and {Perola}, G.~C.},
        title = "{Dynamical Models of Tailed Radio Sources in Clusters of Galaxies}",
      journal = {\aap},
         year = 1973,
        month = aug,
       volume = {26},
        pages = {423},
       adsurl = {https://ui.adsabs.harvard.edu/abs/1973A&A....26..423J}
}

@article{Arshakian2000,
   title={An asymmetric relativistic model for classical double radio sources},
   volume={311},
   ISSN={1365-2966},
   url={http://dx.doi.org/10.1046/j.1365-8711.2000.03098.x},
   DOI={10.1046/j.1365-8711.2000.03098.x},
   number={4},
   journal={Monthly Notices of the Royal Astronomical Society},
   publisher={Oxford University Press (OUP)},
   author={Arshakian, T. G. and Longair, M. S.},
   year={2000},
   month=feb, pages={846–860} }

@INPROCEEDINGS{Lacy2025,
       author = {{Lacy}, Mark and {The VLASS Team}},
        title = "{The VLA Sky Survey}",
    booktitle = {Astronomical Data Analysis Software and Systems XXXIII},
         year = 2025,
       editor = {{Jacques}, Alice and {Seaman}, Robert and {Gandilo}, Natalie and {Linder}, Tyler},
       series = {Astronomical Society of the Pacific Conference Series},
       volume = {541},
        month = oct,
        pages = {416},
          doi = {10.26624/BDYI3779},
       adsurl = {https://ui.adsabs.harvard.edu/abs/2025ASPC..541..416L}
}

@article{Jurlin2020,
	author = {{Jurlin, N.} and {Morganti, R.} and {Brienza, M.} and {Mandal, S.} and {Maddox, N.} and {Duncan, K. J.} and {Shabala, S. S.} and {Hardcastle, M. J.} and {Prandoni, I.} and {R\"ottgering, H. J. A.} and {Mahatma, V.} and {Best, P. N.} and {Mingo, B.} and {Sabater, J.} and {Shimwell, T. W.} and {Tasse, C.}},
	title = {The life cycle of radio galaxies in the LOFAR Lockman Hole field},
	DOI= "10.1051/0004-6361/201936955",
	url= "https://doi.org/10.1051/0004-6361/201936955",
	journal = {A\&A},
	year = 2020,
	volume = 638,
	pages = "A34",
}

@article{de_Villiers2022,
doi = {10.3847/1538-3881/ac460a},
url = {https://doi.org/10.3847/1538-3881/ac460a},
year = {2022},
month = {feb},
publisher = {The American Astronomical Society},
volume = {163},
number = {3},
pages = {135},
author = {de Villiers, Mattieu S. and Cotton, William D.},
title = {MeerKAT Primary-beam Measurements in the L Band},
journal = {The Astronomical Journal}
}

@ARTICLE{Pirya2012,
       author = {{Pirya}, A. and {Saikia}, D.~J. and {Singh}, M. and {Chandola}, H.~C.},
        title = "{A study of the environments of large radio galaxies using SDSS}",
      journal = {\mnras},
         year = 2012,
        month = oct,
       volume = {426},
       number = {1},
        pages = {758-763},
          doi = {10.1111/j.1365-2966.2012.21656.x},
archivePrefix = {arXiv},
       eprint = {1207.1566},
 primaryClass = {astro-ph.CO},
       adsurl = {https://ui.adsabs.harvard.edu/abs/2012MNRAS.426..758P}
}

@ARTICLE{Sethi2025,
       author = {{Sethi}, S. and {Dabhade}, P. and {Biju}, K.~G. and {Stalin}, C.~S. and {Jamrozy}, M.},
        title = "{Study of giant radio galaxies using spectroscopic observations from the Himalayan Chandra Telescope}",
      journal = {\aap},
         year = 2025,
        month = mar,
       volume = {695},
          eid = {A137},
        pages = {A137},
          doi = {10.1051/0004-6361/202553861},
archivePrefix = {arXiv},
       eprint = {2502.06068},
 primaryClass = {astro-ph.GA},
       adsurl = {https://ui.adsabs.harvard.edu/abs/2025A&A...695A.137S}
}

@ARTICLE{Andernach2025,
       author = {{Andernach}, H. and {Br{\"u}ggen}, M.},
        title = "{Properties of giant radio galaxies larger than 3 Mpc}",
      journal = {\aap},
         year = 2025,
        month = jul,
       volume = {699},
          eid = {A257},
        pages = {A257},
          doi = {10.1051/0004-6361/202452961},
archivePrefix = {arXiv},
       eprint = {2505.09181},
 primaryClass = {astro-ph.GA},
       adsurl = {https://ui.adsabs.harvard.edu/abs/2025A&A...699A.257A}
}

@ARTICLE{Lara2004,
       author = {{Lara}, L. and {Giovannini}, G. and {Cotton}, W.~D. and {Feretti}, L. and {Marcaide}, J.~M. and {M{\'a}rquez}, I. and {Venturi}, T.},
        title = "{A new sample of large angular size radio galaxies. III. Statistics and evolution of the grown population}",
      journal = {\aap},
         year = 2004,
        month = jul,
       volume = {421},
        pages = {899-911},
          doi = {10.1051/0004-6361:20035676},
archivePrefix = {arXiv},
       eprint = {astro-ph/0404373},
 primaryClass = {astro-ph},
       adsurl = {https://ui.adsabs.harvard.edu/abs/2004A&A...421..899L}
}

@ARTICLE{McCarthy1991,
       author = {{McCarthy}, Patrick J. and {van Breugel}, Wil and {Kapahi}, Vijay K.},
        title = "{Correlated Radio and Optical Asymmetries in Powerful Radio Sources}",
      journal = {\apj},
         year = 1991,
        month = apr,
       volume = {371},
        pages = {478},
          doi = {10.1086/169911},
       adsurl = {https://ui.adsabs.harvard.edu/abs/1991ApJ...371..478M}
}

@ARTICLE{Ishwara-Chandra1999,
       author = {{Ishwara-Chandra}, C.~H. and {Saikia}, D.~J.},
        title = "{Giant radio sources}",
      journal = {\mnras},
         year = 1999,
        month = oct,
       volume = {309},
       number = {1},
        pages = {100-112},
          doi = {10.1046/j.1365-8711.1999.02835.x},
archivePrefix = {arXiv},
       eprint = {astro-ph/9902252},
 primaryClass = {astro-ph},
       adsurl = {https://ui.adsabs.harvard.edu/abs/1999MNRAS.309..100I}
}

@ARTICLE{Schoenmakers2000,
       author = {{Schoenmakers}, A.~P. and {Mack}, K.-H. and {de Bruyn}, A.~G. and {R{\"o}ttgering}, H.~J.~A. and {Klein}, U. and {van der Laan}, H.},
        title = "{A new sample of giant radio galaxies from the WENSS survey. II. A multi-frequency radio study of a complete sample: Properties of the radio lobes and their environment}",
      journal = {\aaps},
         year = 2000,
        month = oct,
       volume = {146},
        pages = {293-322},
          doi = {10.1051/aas:2000267},
archivePrefix = {arXiv},
       eprint = {astro-ph/0008246},
 primaryClass = {astro-ph},
       adsurl = {https://ui.adsabs.harvard.edu/abs/2000A&AS..146..293S}
}

@ARTICLE{Macklin1981,
       author = {{Macklin}, J.~T.},
        title = "{The symmetry, misalignment and kinematic evolution of double radio sources}",
      journal = {\mnras},
         year = 1981,
        month = sep,
       volume = {196},
        pages = {967-986},
          doi = {10.1093/mnras/196.4.967},
       adsurl = {https://ui.adsabs.harvard.edu/abs/1981MNRAS.196..967M}
}

@ARTICLE{Best1995,
       author = {{Best}, P.~N. and {Bailer}, D.~M. and {Longair}, M.~S. and {Riley}, J.~M.},
        title = "{Radio source asymmetries and unified schemes}",
      journal = {\mnras},
         year = 1995,
        month = aug,
       volume = {275},
       number = {4},
        pages = {1171-1184},
          doi = {10.1093/mnras/275.4.1171},
       adsurl = {https://ui.adsabs.harvard.edu/abs/1995MNRAS.275.1171B}
}

@article{Mahatma2023,
   title={The Dynamics and Energetics of Remnant and Restarting RLAGN},
   volume={11},
   ISSN={2075-4434},
   url={http://dx.doi.org/10.3390/galaxies11030074},
   DOI={10.3390/galaxies11030074},
   number={3},
   journal={Galaxies},
   publisher={MDPI AG},
   author={Mahatma, Vijay H.},
   year={2023},
   month=June, pages={74} }

@article{Morganti2021,
author = {Morganti, Raffaella and Oosterloo, Tom and Murthy, Suma and Tadhunter, Clive},
title = {The impact of young radio jets traced by cold molecular gas},
journal = {Astronomische Nachrichten},
volume = {342},
number = {9-10},
pages = {1135-1139},
doi = {https://doi.org/10.1002/asna.20210037},
url = {https://onlinelibrary.wiley.com/doi/abs/10.1002/asna.20210037},
eprint = {https://onlinelibrary.wiley.com/doi/pdf/10.1002/asna.20210037},
year = {2021}
}

@article{Merritt2002,
   title={Tracing Black Hole Mergers Through Radio Lobe Morphology},
   volume={297},
   ISSN={1095-9203},
   url={http://dx.doi.org/10.1126/science.1074688},
   DOI={10.1126/science.1074688},
   number={5585},
   journal={Science},
   publisher={American Association for the Advancement of Science (AAAS)},
   author={Merritt, David and Ekers, R. D.},
   year={2002},
   month=Aug, pages={1310–1313} }

@ARTICLE{Capetti2002,
       author = {{Capetti}, A. and {Zamfir}, S. and {Rossi}, P. and {Bodo}, G. and {Zanni}, C. and {Massaglia}, S.},
        title = "{On the origin of X-shaped radio-sources: New insights from the properties of their host galaxies}",
      journal = {\aap},
         year = 2002,
        month = oct,
       volume = {394},
        pages = {39-45},
          doi = {10.1051/0004-6361:20021070},
archivePrefix = {arXiv},
       eprint = {astro-ph/0207333},
 primaryClass = {astro-ph},
       adsurl = {https://ui.adsabs.harvard.edu/abs/2002A&A...394...39C}
}

@article{Hardcastle2020,
   title={Radio galaxies and feedback from AGN jets},
   volume={88},
   ISSN={1387-6473},
   url={http://dx.doi.org/10.1016/j.newar.2020.101539},
   DOI={10.1016/j.newar.2020.101539},
   journal={New Astronomy Reviews},
   publisher={Elsevier BV},
   author={Hardcastle, M.J. and Croston, J.H.},
   year={2020},
   month=June, pages={101539} }

@ARTICLE{Mingo2019,
       author = {{Mingo}, B. and {Croston}, J.~H. and {Hardcastle}, M.~J. and {Best}, P.~N. and {Duncan}, K.~J. and {Morganti}, R. and {Rottgering}, H.~J.~A. and {Sabater}, J. and {Shimwell}, T.~W. and {Williams}, W.~L. and {Brienza}, M. and {Gurkan}, G. and {Mahatma}, V.~H. and {Morabito}, L.~K. and {Prandoni}, I. and {Bondi}, M. and {Ineson}, J. and {Mooney}, S.},
        title = "{Revisiting the Fanaroff-Riley dichotomy and radio-galaxy morphology with the LOFAR Two-Metre Sky Survey (LoTSS)}",
      journal = {\mnras},
         year = 2019,
        month = sep,
       volume = {488},
       number = {2},
        pages = {2701-2721},
          doi = {10.1093/mnras/stz1901},
archivePrefix = {arXiv},
       eprint = {1907.03726},
 primaryClass = {astro-ph.GA},
       adsurl = {https://ui.adsabs.harvard.edu/abs/2019MNRAS.488.2701M}
}

@ARTICLE{Mingo2022,
       author = {{Mingo}, B. and {Croston}, J.~H. and {Best}, P.~N. and {Duncan}, K.~J. and {Hardcastle}, M.~J. and {Kondapally}, R. and {Prandoni}, I. and {Sabater}, J. and {Shimwell}, T.~W. and {Williams}, W.~L. and {Baldi}, R.~D. and {Bonato}, M. and {Bondi}, M. and {Dabhade}, P. and {G{\"u}rkan}, G. and {Ineson}, J. and {Magliocchetti}, M. and {Miley}, G. and {Pierce}, J.~C.~S. and {R{\"o}ttgering}, H.~J.~A.},
        title = "{Accretion mode versus radio morphology in the LOFAR Deep Fields}",
      journal = {\mnras},
         year = 2022,
        month = apr,
       volume = {511},
       number = {3},
        pages = {3250-3271},
          doi = {10.1093/mnras/stac140},
archivePrefix = {arXiv},
       eprint = {2201.04433},
 primaryClass = {astro-ph.GA},
       adsurl = {https://ui.adsabs.harvard.edu/abs/2022MNRAS.511.3250M}
}

@ARTICLE{Tadhunter2016,
       author = {{Tadhunter}, Clive},
        title = "{Radio AGN in the local universe: unification, triggering and evolution}",
      journal = {\aapr},
         year = 2016,
        month = jun,
       volume = {24},
       number = {1},
          eid = {10},
        pages = {10},
          doi = {10.1007/s00159-016-0094-x},
archivePrefix = {arXiv},
       eprint = {1605.08773},
 primaryClass = {astro-ph.GA},
       adsurl = {https://ui.adsabs.harvard.edu/abs/2016A&ARv..24...10T}
}

@article{Lal2004,
author = {Lal, Dharam and Rao, Aroor},
year = {2004},
month = {12},
pages = {},
title = {Spectral structure of X-shaped radio sources},
journal = {Bulletin of the Astronomical Society of India}
}

@article{Lal2006,
author = {Lal, Dharam and Rao, Aroor},
year = {2006},
month = {11},
pages = {},
title = {GMRT observations of X-shaped radio sources}
}

@ARTICLE{Riley1972,
       author = {{Riley}, J.~M.},
        title = "{Observations of 3C 272.1 at 2.7 and 5.0 GHz}",
      journal = {\mnras},
         year = 1972,
        month = jan,
       volume = {157},
        pages = {349-375},
          doi = {10.1093/mnras/157.4.349},
       adsurl = {https://ui.adsabs.harvard.edu/abs/1972MNRAS.157..349R}
}

@book{Pacholczyk1970,
  author = {Pacholczyk, A. G.},
  title = {Radio Astrophysics: Nonthermal Processes in Galactic and Extragalactic Sources},
  publisher = {W. H. Freeman},
  address = {San Francisco},
  adsurl = {https://ui.adsabs.harvard.edu/abs/1970ranp.book.....P},
  year = {1970}
}

@ARTICLE{Tribble1993,
       author = {{Tribble}, Peter C.},
        title = "{Radio spectral ageing in a random magnetic field.}",
      journal = {\mnras},
         year = 1993,
        month = mar,
       volume = {261},
        pages = {57-62},
          doi = {10.1093/mnras/261.1.57},
       adsurl = {https://ui.adsabs.harvard.edu/abs/1993MNRAS.261...57T}
}
\bibliographystyle{aasjournalv7}

\end{document}